\documentclass[11pt]{article}
\usepackage[utf8]{inputenc}
\usepackage{setspace}
\usepackage{geometry}
\usepackage{array}
\usepackage{booktabs}
\usepackage{hyperref}
\usepackage{graphicx}
\usepackage{url}
\usepackage{xcolor, colortbl}
\usepackage{enumitem}
\usepackage{authblk}
\usepackage{amssymb}
\usepackage{algorithm}
\usepackage{multirow}
\usepackage{longtable}

\usepackage[numbers]{natbib}

\urldef{\mailsa}\path|ricardo.fitas@tu-darmstadt.de,|

\newcommand{\keywords}[1]{%
  \par\addvspace\baselineskip
  \noindent\textbf{Keywords:}\enspace\ignorespaces#1
}

\begin{document}

\title{Inclusive Education with AI: Supporting Special Needs and Tackling Language Barriers}

\author{
    Ricardo Fitas \\
    Technical University of Darmstadt, Darmstadt, Hesse, Germany \\
    \texttt{ricardo.fitas@tu-darmstadt.de}\\
    \href{https://orcid.org/0000-0001-5137-2451}{https://orcid.org/0000-0001-5137-2451}
}

\date{}
\maketitle

\begin{abstract}
\emph{Early childhood classrooms are becoming increasingly diverse, with students spanning a range of linguistic backgrounds and abilities. AI offers innovative tools to help educators create more inclusive learning environments by breaking down language barriers and providing tailored support for children with special needs. This chapter provides a comprehensive review of how AI technologies can facilitate inclusion in early education. It is discussed AI-driven language assistance tools that enable real-time translation and communication in multilingual classrooms, and it is explored assistive technologies powered by AI that personalize learning for students with disabilities. The implications of these technologies for teachers are examined, including shifts in educator roles and workloads. General outcomes observed with AI integration---such as improved student engagement and performance---as well as challenges related to equitable access and the need for ethical implementation are highlighted. Finally, practical recommendations for educators, policymakers, and developers are offered to collaboratively harness AI in a responsible manner, ensuring that its benefits reach all learners.}
\end{abstract}


\section{Introduction}
Inclusive education aims to ensure that all learners, including those with disabilities or language barriers, have equitable access to quality learning \cite{dudu2024inclusive,NavyaPoly2023}. Yet globally, challenges remain immense: an estimated 240~million children live with disabilities, and about half of them are out of school entirely, while those in school often lack adequate support \cite{InternationalCommission2021}. Likewise, roughly 40\% of the worlds population does not receive education in a language they speak or understand \cite{UNESCO2022multilingual}, highlighting pervasive language barriers. These gaps underscore the urgency for innovative solutions. Artificial intelligence (AI) has emerged as a promising tool to address such disparities by providing personalized support and adaptive learning resources at scale \cite{OluwaseyiAinaGboladeOpesemowo2024,hongli2024ai}. International organizations have emphasized that AI, if leveraged properly, can "\textit{strengthen inclusion, improve the quality of learningand expand access to knowledge,} \cite{UNESCO2023AI, OECD2021} though they caution that realizing this potential requires ethical and inclusive implementation frameworks. Early childhood education, in particular, may greatly benefit from AI-driven assistive technologies  young learners with diverse needs can engage with AI-powered educational games \cite{MehdiRoopaei2024}, translation apps \cite{MuhammadUsmanTariq2024}, and personalized tutors \cite{GSornavalli2024} that adapt to their developmental level, potentially improving learning outcomes and inclusion from the start \cite{SalasPilco2022, HarnessingAI2024}. Educators can better accommodate each childs linguistic and special needs by integrating AI into the early learning environment \cite{Kaimara2023, AIinSpecialEd2022}.

The integration of AI in special education and language learning presents both promising opportunities and significant challenges. While AI-based interventions have shown potential in enhancing learning outcomes for students with disabilities, the effects are often modest and lack longitudinal evidence of sustained benefits. For instance, a meta-analysis of AI interventions for students with disabilities revealed a medium effect size, indicating some positive impact but underscoring the need for more rigorous research to ensure accessibility and agency for these students \cite{Zhang2024}. Similarly, AI applications for students with Autism Spectrum Disorder (ASD) have demonstrated potential in supporting learning, yet gaps remain in understanding long-term effectiveness and standardizing methodologies \cite{kotsi2025}. AI's role in addressing language impairments and Language-Based Learning Disorders (LBLDs) is promising, offering personalized and adaptive learning experiences, but requires careful implementation to avoid over-reliance on technology and ensure equitable access \cite{mahmoudi-dehaki2024, mohebbi2024}. The transformative potential of AI in language education is evident, with significant improvements in language proficiency reported, yet future research must focus on long-term effects and ethical implementation to foster inclusive environments \cite{alzahrani2024}. Despite these advancements, disparities in AI adoption between wealthier and poorer schools risk creating digital divides, undermining inclusion goals \cite{modi2024}. Game-based AI has shown effectiveness in improving literacy skills and engagement for students with Specific Learning Disorders, suggesting innovative solutions for differentiated learning \cite{sukasih2024}. Machine learning models have proven effective in early detection of learning disabilities, facilitating personalized interventions and enhancing diagnostic accuracy \cite{jain2024}. AI-based adaptive learning has significantly improved academic performance and social interaction for students with special needs, highlighting its potential in inclusive education settings \cite{maulidin2024}. A meta-analysis of AI in education further supports its large effect on learning achievement, particularly through chatbots and personalized learning systems, but emphasizes the need for understanding moderating factors such as intervention duration and geographical distribution \cite{tlili2025}. Collectively, these studies highlight the need for a careful scholarly examination of AI's role in inclusive education, addressing research gaps in efficacy, bias, and access to ensure AI benefits all learners, not just a privileged few.

The objectives of this chapter are to critically explore the role of AI in fostering inclusive education during early learning, specifically by supporting students with special needs and those facing language barriers. The author targets an audience of educational researchers, practitioners (teachers, school leaders), and policymakers interested in leveraging emerging technologies for equity in education. Through an integrated review of recent literature and case studies, this paper aims to:

\begin{itemize}
\item identify how AI tools are being applied to assist with language translation and learning in diverse classrooms;

\item examine AI-driven assistive technologies designed for students with disabilities and their documented impacts;

\item analyze the influence of these AI innovations on teachers roles and workloads;

\item discuss overall outcomes, challenges, and best practices for ethical integration of AI in inclusive settings. 

\end{itemize}

The chapter aims to give educators and decision-makers practical insights and suggestions to help them embrace AI in ways that foster inclusion rather than worsen inequality by combining findings from several fields.

The remainder of this chapter is as follows: Section~2 details the data sources and selection of those that are included in the current research. Section~3 examines AI solutions for language assistance and translation in education, including tools and case studies that address linguistic barriers. Section~4 discusses AI-driven assistive technologies for special-needs students, with examples of personalized learning supports and their real-world impacts. Section~5 considers how AI integration influences early educators roles and workloads, analyzing ways AI can reduce routine tasks and enable more personalized teaching, alongside implementation challenges and ethical considerations in teacher-AI dynamics. Section~6 then outlines general outcomes observed from AI-inclusive education initiatives  highlighting positive impacts on engagement and learning, as well as persistent challenges. Building on these findings, Section~7 provides recommendations for ethical AI integration, including strategic, policy, and research directions to ensure AI tools are used responsibly and effectively to support all learners. Finally, Section~8 concludes the chapter with a summary of key insights, a call to action for stakeholders, and suggestions for future research to continue advancing inclusive education with AI.

\section{Data Sources and Selection}
This chapter is based on a comprehensive review of literature and reports at the intersection of AI, early childhood education, inclusion, and accessibility. The authors surveyed peer-reviewed journals, conference proceedings, and academic databases (e.g., Google Scholar, IEEE Xplore, Springer, ERIC) for articles published primarily in the last 3 years. In addition, a selection of authoritative white papers and policy reports from government and international organizations to capture practical insights and guidelines were also included. Key search terms included combinations of inclusive education, artificial intelligence, special needs, assistive technology, early childhood, language barrier, and translation tools, among others. To ensure relevance, the authors applied the following inclusion criteria: the study or article explicitly addresses AI (broadly including machine learning, educational software intelligence, etc.) in an educational context; it involves learners with disabilities and/or multilingual settings; and it discusses outcomes, benefits, or challenges related to inclusion. Sources that only tangentially mentioned AI in education or lacked any implications for education or teaching were excluded. Using these criteria, the authors gathered a diverse yet focused dataset of publications that inform the subsequent sections. 

\section{AI for Language Assistance and Translation}
\subsection{Language Barriers \& AI Solutions}
Language differences in the classroom can significantly hinder a students ability to learn and a teachers ability to include everyone \cite{mokikwa2024navigating}. In many regions, students speaking minority languages or recent immigrants learning the school language struggle to keep up with instruction delivered only in the dominant tongue \cite{CansuYILDIZ2024,SadiaHameed2025}. Traditionally, schools have addressed this through bilingual aides, translated materials, or peer support, but such resources are often limited \cite{Holmes2022,UNESCO2024multilingual}. AI offers new scalable solutions to bridge these language gaps. Machine translation systems, powered by advanced neural networks, can now instantly translate text and speech between dozens of languages with reasonably high accuracy \cite{SEAMLESS2025}. In an educational context, this means an English-speaking teacher can communicate basic instructions to, say, a Ukrainian-speaking or Arabic-speaking student via a tablet or smartphone app that translates and voices the content in the students native language. Conversely, the student could respond in their language and have it translated to English for the teacher. Speech recognition combined with translation can produce real-time captions of a teachers lesson in multiple languages, allowing learners to follow along by reading in a language they understand \cite{USDeptEd2023}. These AI-driven solutions directly target one of the biggest inclusion challenges worldwide  as noted earlier, millions of learners are currently taught in languages they do not fully comprehend, suppressing their learning potential \cite{UNESCO2024multilingual,Badawi2024}. Instead of expecting children to be able to adapt to a foreign language environment, schools can use AI language technologies to adjust to their linguistic needs dynamically.

Natural language processing (NLP) innovations have led to a variety of tools that can support multilingual education. For example, large language models (such as GPT-based systems) can simplify complex text or generate bilingual glossaries for lesson content, helping second-language learners grasp key concepts \cite{su17030956}. AI-driven lexical tools and reading support systems can deliver instant explanations or translations for challenging vocabulary encountered by learners during their reading activities. These AI-driven tools exemplify the principle of enhanced accessibility, allowing students to obtain an additional layer of understanding or translation that goes beyond the conventional resources available to them. Importantly, AI solutions are not limited to the major world languages; there are ongoing efforts to train translation models for less-resourced languages and dialects, which could be transformative for minority language communities \cite{Benboujja2024,UNESCO2024multilingual}. For instance, a project in Peru developed the \textit{PictoAndes} system  an AI-enhanced communication board that supports Spanish and indigenous languages  to assist children with communication difficulties in multilingual settings \cite{su17030956}. While originally aimed at special needs, such innovations also demonstrate how AI can serve multicultural inclusion by handling languages that teachers may not know \cite{Holmes2022}.

\paragraph{Summary and Classroom Implications.}
AI-powered translation tools can significantly enhance classroom inclusivity by supporting learners from diverse linguistic backgrounds. In real classroom settings, educators can integrate real-time translation apps, such as Google Translate or DeepL, to assist with instructional communication. These tools reduce delays in comprehension and help foster a more collaborative and engaged learning atmosphere.

\subsection{Translation Tools in the Classroom}
\begin{figure}
    \centering
    \includegraphics[width=\linewidth]{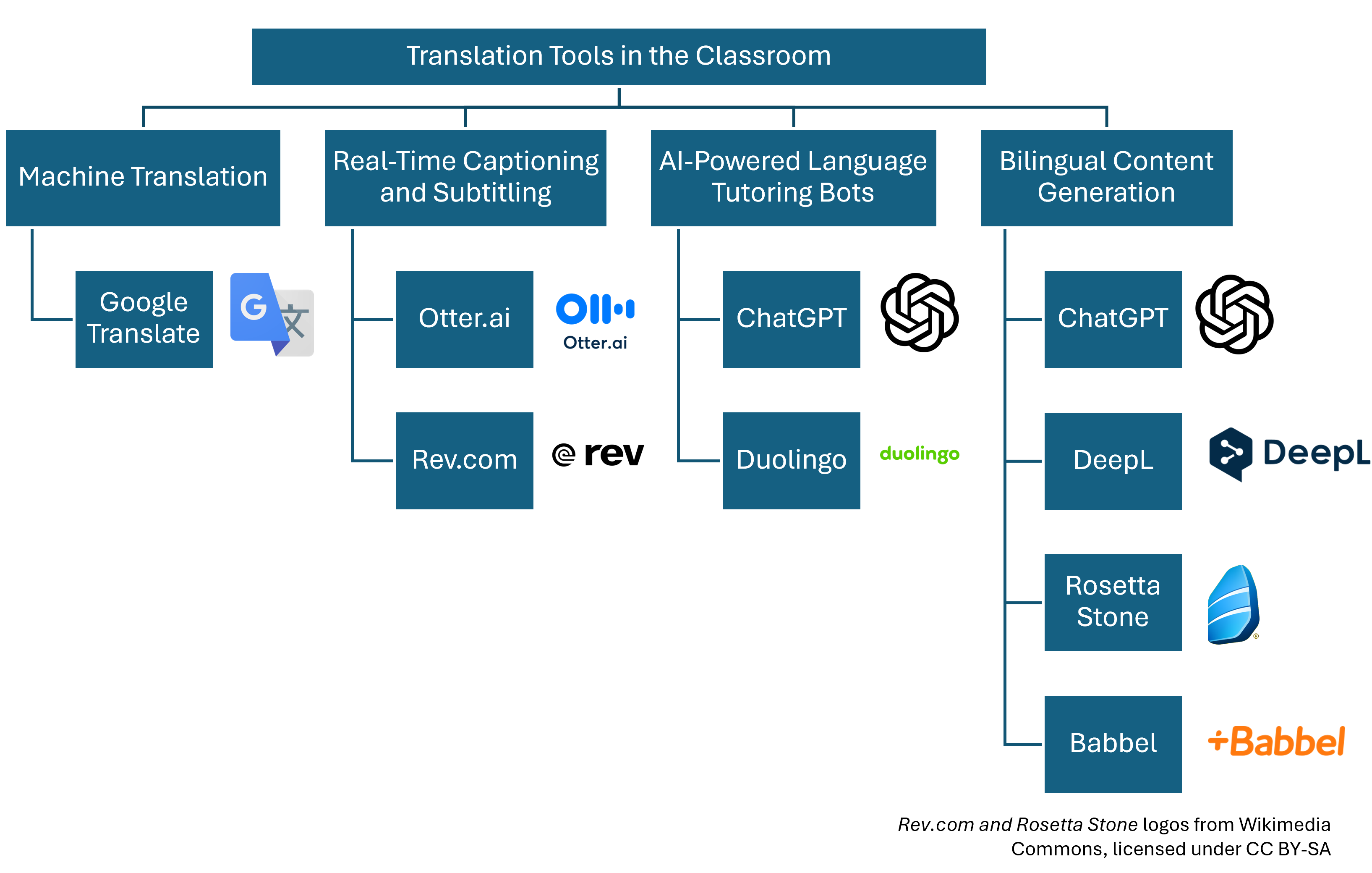}
    \caption{Visualization of the AI translation tools in the classroom}
    \label{fig1}
\end{figure}

In practice, several AI translation tools have already made their way into classrooms around the world \cite{DAntonyArulRaj2024}. Figure \ref{fig1} summarizes the mentioned AI translation tools in this section.

\subsubsection{Machine Translation Apps}

Machine translation apps are among the most widely used AI tools in language education. These tools leverage advanced algorithms to translate text and speech in real-time, facilitating communication across languages.

\begin{itemize}

\item{Google Translate}

Google Translate is one of the most popular machine translation tools, widely used in educational settings. It supports over 100 languages and offers features such as text translation, speech translation, and instant camera translation. For instance, ESL (English as a Second Language) students often use Google Translate to translate large amounts of text related to their subjects, while teachers can use it to provide new tools and opportunities for language instruction \cite{Shevchenko2023, WaelAlharbi2023}. The app's advanced features, such as machine learning, have improved its efficiency and versatility, making it an indispensable resource for both learners and educators.

\item{AI-Powered Translation Tools in Multilingual Classrooms}

AI-powered translation tools are increasingly being deployed in multilingual classrooms to improve accessibility for non-native speakers. These tools not only facilitate communication between students and teachers but also foster an inclusive learning environment. For example, real-world case studies have demonstrated the effectiveness of AI translation platforms in breaking down linguistic barriers and promoting cross-cultural understanding \cite{MuhammadUsmanTariq2024, TonyNkomoki2025}.

\end{itemize}

\subsubsection{Real-Time Captioning and Subtitling}

Real-time captioning and subtitling are essential for creating accessible learning environments, particularly for students with hearing impairments or those learning a new language.

\begin{itemize}

\item{AI-Driven Captioning Tools}

AI-driven tools like Otter.ai and Rev.com provide real-time captioning, enabling students to follow lectures and discussions more effectively. These tools are particularly beneficial in language learning contexts, where learners can see spoken language in text form, helping them improve their listening and comprehension skills. Research has shown that such tools enhance learner engagement and provide personalized learning experiences \cite{AliaSaadEldinAbusahyon2023, AbduvalievaDilnozaIsmoilovna2025}.

\item{Subtitling in Language Learning}

Subtitling is another AI-powered feature that supports language acquisition. For instance, platforms like YouTube and Netflix offer AI-generated subtitles in multiple languages, allowing learners to practice listening and reading skills simultaneously. This feature is particularly useful for language learners who can toggle between subtitles in their native language and the target language to better understand spoken content \cite{DanaKristiawan2024, HafizaSaumiRamadilla2025}.

\end{itemize}

\subsubsection{AI-Powered Language Tutoring Bots}

AI-powered tutoring bots are revolutionizing language learning by providing personalized and interactive learning experiences.

\begin{itemize}

\item{ChatGPT and Language Learning}

ChatGPT, an AI-powered chatbot, has emerged as a powerful tool for language learning. It offers customized learning experiences based on the learner's goals and proficiency level. Through chat-based interactions, learners can practice grammar, vocabulary, and pronunciation, receiving immediate feedback to correct mistakes in real-time. ChatGPT also simulates real-world situations, allowing learners to practice their language skills in practical contexts \cite{MahshadDavoodifard2024, Shevchenko2023, WeimingLiu2024}.

\item{Duolingo and AI-Powered Tutors}

Duolingo, a popular language learning app, integrates AI-powered tutors to provide personalized feedback and adaptive learning pathways. The app uses natural language processing (NLP) and machine learning algorithms to assess learners' progress and tailor instruction accordingly. This approach has been shown to enhance learner autonomy and engagement, making language learning more accessible and enjoyable \cite{BorhadeRani2024, MichailStFountoulakis2024}.

\end{itemize}

\subsubsection{Bilingual Content Generation}

Bilingual content generation is another innovative application of AI in language education, enabling the creation of educational materials in multiple languages.

\begin{itemize}

\item{Generative AI and Language Learning}

Generative AI tools, such as ChatGPT and DeepL, can generate bilingual content, including texts, exercises, and quizzes, tailored to specific learning needs. These tools are particularly useful for creating personalized learning materials for students with diverse linguistic backgrounds. For example, teachers can use generative AI to develop bilingual resources that cater to the needs of multilingual classrooms, promoting inclusivity and equity in education \cite{MahshadDavoodifard2024, RoleofArtificialIntelligenceinFacilitatingEnglishLanguageLearningforNonNativeSpeakers2024}.

\item{AI-Driven Language Learning Platforms}

AI-driven platforms like Rosetta Stone and Babbel use generative AI to create bilingual content, offering interactive modules that simulate real-life language use. These platforms provide learners with opportunities to practice speaking, listening, reading, and writing skills in an immersive and engaging manner. By leveraging generative AI, these platforms can adapt to individual learner needs, offering a more effective and efficient language learning experience \cite{AbduvalievaDilnozaIsmoilovna2025, LazzatKonyrova2024}.

\end{itemize}

\paragraph{Summary and Classroom Implications.}
By embedding AI translation tools like real-time captioning and subtitling into lesson delivery, educators can ensure students with hearing impairments or second-language learners follow instruction effectively. These technologies increase comprehension and confidence, allowing students to focus more on content mastery than language decoding.

\subsection{Case Studies \& Challenges}
Early case studies of AI translation in education demonstrate both its promise and the challenges to be addressed. In different pilot programs at various U.S. school districts with a large population of multilingual and refugee students, handheld AI translation devices were provided to newcomer students. Over time, teachers observed that these devices helped students better understand lessons and participate more actively in group activities by enabling them to communicate more effectively \cite{pocketalk2023,edtech2024,mtvernon2024}. 

In Egypt and Lebanon, AI-powered opportunistic learning platforms have enabled students to access high-quality educational resources that are culturally and linguistically relevant. These platforms help overcome challenges related to non-native instructional languages by offering adaptive content delivery based on students' proficiency, preferences, and regional context. They also integrate Arabic-English bridging mechanisms through NLP and real-time feedback loops, enhancing cross-cultural communication and learner engagement \cite{hamed2024ai}.

Moreover, across Sub-Saharan Africa, AI tools are being used to personalize learning for students with special needs by adapting materials, feedback, and assessment strategies. One notable example is the deployment of AI-based assistive tutoring and early-intervention platforms that dynamically adjust based on student behavior and progress. These tools not only promote independence and inclusivity but also address digital exclusion by enabling accessibility in linguistically diverse and resource-constrained environments \cite{chisom12023review}.

Researchers \cite{MuhammadAbdeePrajaMukti2024} emphasize the importance of using AI translation tools as a complement to traditional language learning methods rather than as a substitute. This balanced approach is crucial to prevent overreliance on AI, which can impede the development of independent language skills. Although AI tools are beneficial for efficiency and comprehension, there is a risk that students may become overly dependent on them, potentially delaying the development of their language proficiency. The study suggests that AI should be used to support language learning, ensuring that students continue to practice and develop their language skills independently. In addition, the necessity of human oversight is emphasized to ensure translations are accurate and contextually relevant. This oversight is essential to maximize the benefits of AI tools while mitigating their limitations, thus supporting effective language acquisition \cite{MuhammadAbdeePrajaMukti2024}. 

Researchers also underscore the importance of using AI to bridge language barriers effectively, focusing on accuracy and cultural relevance \cite{benboujja2024overcoming}. However, to prevent potential distractions in a classroom environment, educators should develop strategies that dictate when and how these tools are used. For instance, AI translators could be employed during independent study sessions to aid comprehension, while encouraging students to engage actively without translation aids during interactive discussions or storytelling sessions to foster listening skills and real-time engagement. 

Another issue is the accuracy and bias of AI language tools. While major languages are well-supported, students speaking less common languages or dialects may find the AI struggles to translate accurately, which could lead to misunderstandings or frustration. Moreover, idiomatic expressions or context-dependent meanings can be mistranslated; for instance, an AI might translate a phrase literally and inadvertently convey the wrong meaning in the target language \cite{UNESCO2024multilingual}. Such errors not only confuse students but could even cause social embarrassment or offense.

There is also the question of equitable access \cite{SagafSPettalongi2024,YalalemAssefa2024,VivekAhuja2023,FarhatNaureenMemon2024}, and not all schools can afford individual devices or have reliable internet for cloud-based translation services \cite{OECD2021,USDeptEd2023}. Without careful planning, the introduction of AI for language support could widen disparities between well-resourced and under-resourced schools (this concern is discussed further in Section~6). Finally, some teachers worry that over-reliance on translation technology might impede students motivation to learn the mainstream language, similar to how overusing a calculator might hinder mastering arithmetic. These challenges underscore that AI language assistance must be implemented thoughtfully: it works best as an aid under teacher guidance, combined with intentional strategies to eventually wean students off constant translation as they gain proficiency.

\subsubsection*{Best Practice Steps}
\begin{enumerate}
    \item Select a translation tool that supports the students native language.
    \item Train students briefly on how to use the tool.
    \item Set classroom norms for when to use AI translation versus direct communication.
    \item Monitor usage and encourage gradual reduction as proficiency increases.
\end{enumerate}

\paragraph{Summary and Classroom Implications.}
AI language tools, while powerful, must be implemented strategically. Teachers should balance their use with traditional teaching to promote long-term language acquisition. In class, AI should complement instructionfor example, supporting vocabulary during reading time but minimized during collaborative dialogue to promote fluency development.

\subsection{Benefits \& Limitations}

When used appropriately, AI-driven language assistance can yield significant benefits for inclusion. Students who once felt lost due to language barriers gain a sense of belonging and confidence when they can understand classroom discussions and express themselves with AI support \cite{Benboujja2024}. This improved participation often leads to better academic progress; for example, English learners provided with real-time translation and bilingual content have been shown to perform closer to their native-speaking peers in comprehension assessments \cite{article23gz,wei2023artificial}.

Additionally, these tools can serve as accelerators for language learningby providing immediate translations, they create countless teachable moments where students pick up new vocabulary and phrases in the new language \cite{BarnoBerdiyevaTurdaliyevna2024,AhmadrezaMohebbi2024}. Over time, many students begin to rely less on the tool as their proficiency grows, using it only for occasional help \cite{JamesEdwardLalira2024}. Ultimately, these AI-driven translation tools tend to be used more for academic purposes rather than general communication \cite{TonyNkomoki2025}. However, for that student who tends to be highly dependent on AI for continued translation (generally because they cannot develop the necessary skills to start getting fluency), a hybrid approach where traditional methods \cite{UmarUmar2024,PeterCAbernathy2024} and educator involvement \cite{MichailStFountoulakis2024} is generally recommended. Teachers also benefit: instead of spending disproportionate time trying to individually tutor non-native speakers or translate materials, they can offload some of this work to AI and focus on higher-level teaching or one-on-one support that AI cannot provide in order to turn the learning process faster \cite{DilnozaIsmoilovnaAbduvalieva2025,TonyNkomoki2025}.

Culturally, embracing multiple languages via AI sends a message of validation to students identities, aligning with inclusive education principles. Parents who do not speak the school language can also be more involved since communication (notes, meetings, report cards) can be translated for them; this strengthens the home-school partnership for the students benefit \cite{DwiMariyono2024}.

However, there are limitations to what AI language tools can achieve, and recognizing these is important for setting realistic expectations. First, translation AI is not infallible  misunderstandings can arise if students or teachers take AI outputs at face value without verification. Critical information (like safety instructions, medical information, or assessment questions) generally still needs human-translated accuracy \cite{RoyaShahmerdanova2025}.

AI currently also has a limited ability to handle nuanced classroom discourse; humor, sarcasm, or idiomatic classroom banter may not translate well, and the social aspect of communication (body language, tone) is largely lost when mediated through an app \cite{PictoAndes2025}. This means that a student using an AI translator might miss out on some subtleties of interaction, potentially impacting social inclusion \cite{XinyueLi2025,RoyaShahmerdanova2025}.

Moreover, technical requirements can be a limitation: reliable use of cloud-based translation often needs a strong internet connection and modern devices, which may not be available in all schools or regions \cite{DavidCHill2022,zhang2024nya}. Offline models exist for some languages but are less accurate \cite{wang2024optimizing}. There are also concerns about data privacy  using cloud translation means student speech or text is sent to third-party servers (e.g., big tech companies), raising questions about compliance with student data protection laws and whether sensitive information might be inadvertently shared \cite{liu2024pptif}.

From a pedagogical standpoint, teachers must ensure that AI support complements rather than replaces human scaffolding. The goal should be for students to progress in both content mastery and language development; thus, while AI can lighten the linguistic load in the short term, educators should gradually adjust its use to encourage more direct engagement with the new language (for example, by shifting from full-sentence translation to only keyword translation over time) \cite{EyvindElstad2024}. In this sense, the Artificial Intelligence Pedagogical Content Knowledge (AIPACK) framework \cite{NuriBalta2024} also emphasizes the importance of context-sensitive applications of AI, ensuring that its use aligns with specific pedagogical goals and supports rather than replaces human cognitive effort.

In summary, AI-based language assistance is a powerful enabler of inclusive education when its use is balanced with effective teaching strategies and when infrastructure and oversight are in place. It can break down language walls in the classroom, but it is not a magic solution  success depends on addressing technical, pedagogical, and ethical limitations through thoughtful implementation.

\paragraph{Summary and Classroom Implications.}
AI-driven language assistance enhances both comprehension and student participation, especially when integrated into curriculum-aligned activities. However, teachers must remain aware of potential over-dependence, device access inequality, and the ethical issues associated with data collection. Classroom practices should therefore include planned transitions from AI reliance toward autonomous communication.

\paragraph{Section Summary:} Section 3 demonstrated how language-focused AI tools can promote multilingual inclusion, improve student participation, and reduce instructional inequality. Practical tools and case studies helped establish realistic implementation strategies that support both students and educators.

\section{AI and Assistive Technology for Special Needs}

\begin{figure}
    \centering
    \includegraphics[width=0.5\linewidth]{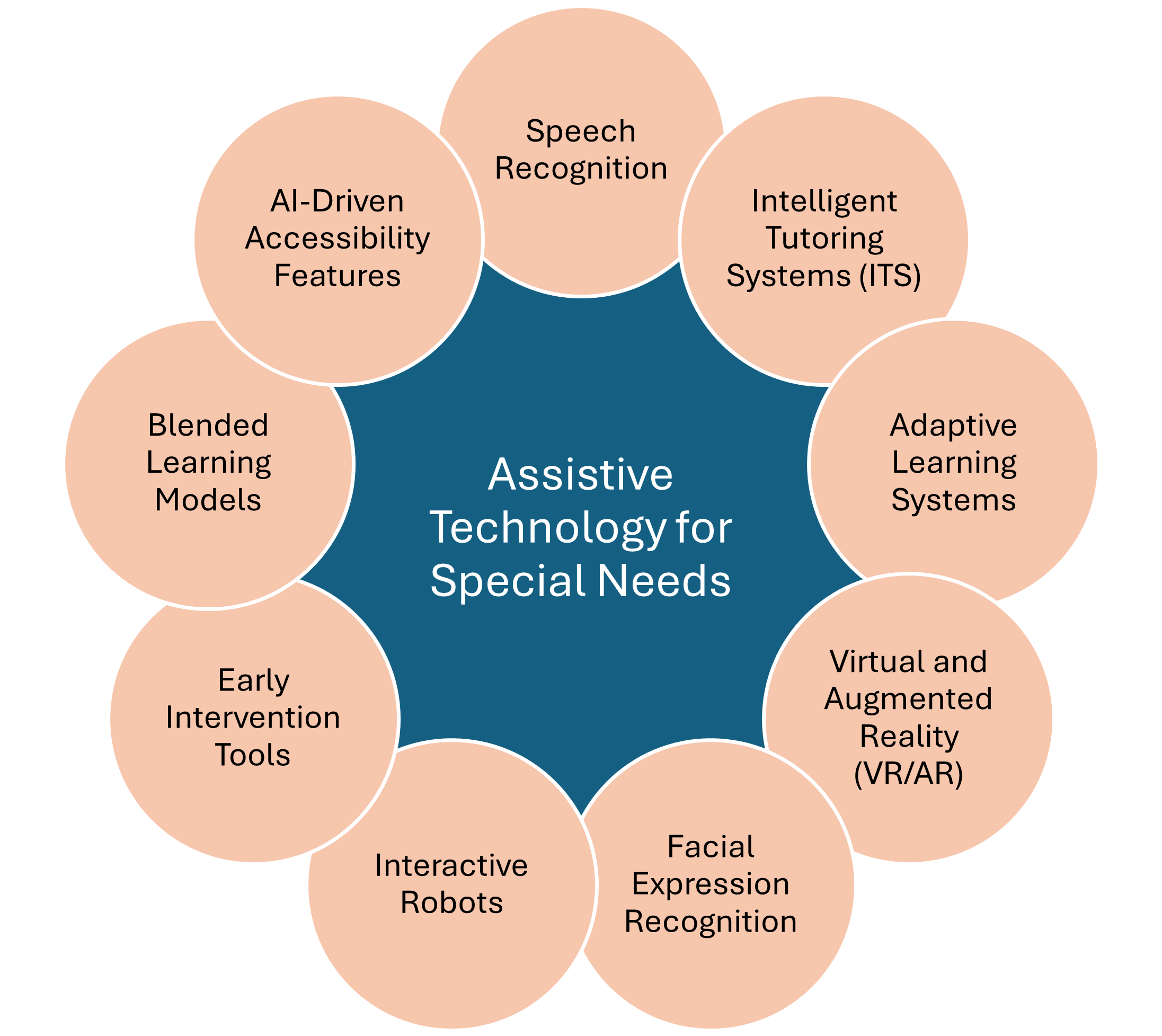}
    \caption{Visualization of all the mentioned categories of assistive technology for special needs}
    \label{fig2}
\end{figure}

The integration of AI in special needs education has revolutionized the way educators support students with diverse learning requirements. AI-driven assistive technologies are tailored to address specific challenges, enhance accessibility, and provide personalized learning experiences. This section categorizes these technologies by type and explores their educational applications, supported by insights from relevant research. Figure \ref{fig2} summarizes all the mentioned categories of assistive technology for special needs. Table \ref{tab:aiassistivetech} further details on the educational application of those categories.

\begin{table}[h]
    \centering
    \caption{Categorization of AI-Driven Assistive Technologies for Special Needs Education}
    \renewcommand{\arraystretch}{1.3}
    \begin{tabular}{p{4cm} p{7cm} p{4cm}}
        \toprule
        \textbf{Technology Type} & \textbf{Educational Application} & \textbf{Examples (with References)} \\
        \midrule
        Intelligent Tutoring Systems (ITS) & Personalized learning pathways for students with learning disabilities & ALEKS, Q-interactive \cite{WasimAhmad2024} \\
        Speech Recognition Tools & Facilitating communication and accessibility for students with speech or hearing impairments & AAC devices, Stamurai \cite{ShivaniShivani2024, ErenKamber2025} \\
        Adaptive Learning Systems & Tailoring educational content to individual learning needs & Carnegie Learning, Notebook \cite{WasimAhmad2024, ShivaniShivani2024} \\
        Virtual and Augmented Reality (VR/AR) & Creating immersive learning environments for students with ASD and other disabilities & Google Glass, Augmenta11y app \cite{WasimAhmad2024, ShivaniShivani2024} \\
        Facial Expression Recognition & Identifying emotional cues to support students with social communication challenges & Chat robots, facial analysis tools \cite{SahrishPanjwaniCharania2024, ToluwaniVictorALIU2024} \\
        Interactive Robots & Providing interactive learning experiences for students with neurodevelopmental disorders & AACDD, DYSXA apps \cite{ShivaniShivani2024, PrabalDattaBarua2022} \\
        Early Intervention Tools & Detecting learning disabilities and developmental delays early & AI-driven diagnostic platforms \cite{SahrishPanjwaniCharania2024, PradnyaMehta2023} \\
        Blended Learning Models & Combining traditional and AI-driven methods for inclusive education & Integration with UDL frameworks \cite{EsmaeilZaraiiZavaraki2024, SYARIFMAULIDIN2024} \\
        AI-Driven Accessibility Features & Enhancing independence and inclusivity for students with disabilities & Braille devices, automated captioning systems \cite{ShivaniShivani2024, ErenKamber2025} \\
        \bottomrule
    \end{tabular}
    
    \label{tab:aiassistivetech}
\end{table}

\subsection{Intelligent Tutoring Systems}

Intelligent Tutoring Systems (ITS) are AI-powered educational tools that provide personalized instruction by adapting to the individual learning pace and style of each student, offering real-time feedback and tailored resources. These systems are particularly beneficial for students with learning disabilities, such as dyslexia and dyscalculia, by creating customized learning pathways and interactive exercises that cater to their specific needs \cite{MMeenakshi2024, SMahalakshmi2025}. ITS platforms like ALEKS and Q-interactive exemplify this by adjusting the difficulty of learning materials based on student responses, ensuring that the content remains appropriately challenging and engaging \cite{GhanchiKabirSarfaraj2025, SAbinaya2024}. The integration of AI in ITS enhances their capabilities, allowing for dynamic adaptation and optimized learning outcomes through the analysis of vast amounts of data to tailor educational experiences to each learner's unique needs \cite{DiptaGomes2024, EmmanuelDumbuya2024}. AI technologies such as natural language processing (NLP), machine learning, and adaptive learning platforms are instrumental in this process, enabling ITS to provide real-time answers, explanations, and formative assessments that are aligned with individual learning patterns and preferences \cite{GhanchiKabirSarfaraj2025, PRathika2024}. Moreover, AI-driven ITS can significantly improve accessibility and engagement for disabled students, offering personalized assistive tools that facilitate learning and foster independence \cite{SMahalakshmi2025, MBhakiyasri2024}. The ethical considerations of AI in education, including privacy and bias mitigation, are crucial to ensure that these systems are implemented effectively and equitably \cite{MeltemTakn2025}. Overall, ITS represent a significant advancement in personalized education, with the potential to revolutionize traditional learning paradigms by providing inclusive, adaptive, and data-driven educational experiences \cite{GSornavalli2024, MeltemTakn2025}.

\paragraph{Summary and Classroom Implications.}
ITS platforms allow teachers to offer customized pathways for students with specific learning disabilities. In the classroom, these tools free educators from one-size-fits-all instruction, supporting differentiation and real-time feedback tailored to each student's needs.

\subsection{Speech Recognition Tools}

Speech recognition and synthesis tools, particularly Speech-to-Text (STT) and Text-to-Speech (TTS) technologies, play a crucial role in enhancing educational accessibility for students with disabilities. These technologies convert spoken language into written text and vice versa, thereby facilitating communication and learning for students with hearing, speech, visual, or reading impairments \cite{DNikolaeva2023, Harinis2024, LifaSun2023}. The integration of TTS in educational settings allows for the conversion of text into natural-sounding speech. This is particularly beneficial for visually impaired students or those with reading difficulties, as it enhances content consumption and comprehension \cite{Harinis2024, YimingWang2024}. The advancements in deep learning have significantly improved the naturalness and intelligibility of TTS systems, making them more effective for diverse educational applications \cite{LifaSun2023}. STT technologies enable the transcription of spoken language into text, which is invaluable for students with speech impairments, allowing them to interact more effectively with educational content \cite{VMReddy2023, SowjanyaJindam2024}. AI-powered tools, such as Stamurai, provide affordable speech therapy support, demonstrating the potential of these technologies to offer personalized and accessible educational experiences \cite{SandraRaffoul2023}. Augmentative and Alternative Communication (AAC) devices and automated captioning systems are integral to inclusive education, ensuring that all students can engage with classroom materials \cite{TundeToyeseOyedokun2024}. The integration of AI-powered solutions into AAC technologies further enhances their effectiveness, fostering independent learning and personal growth \cite{MohammadMuzammilKhan2021}. Despite the significant advancements in STT and TTS, challenges such as transcription inaccuracies, handling diverse accents, and ethical concerns regarding AI-generated content remain \cite{SimonaTomkova2024, VMReddy2023}. Ongoing research is necessary to enhance these technologies' robustness and applicability, ensuring their effectiveness in diverse educational contexts. Overall, the deployment of STT and TTS technologies in education not only supports students with disabilities but also promotes a universal design for learning, fostering independent learning and personal growth \cite{TundeToyeseOyedokun2024, MohammadMuzammilKhan2021}.

\paragraph{Summary and Classroom Implications.}
Speech-to-text and text-to-speech tools enhance accessibility for students with speech or hearing impairments. Teachers can incorporate these tools into reading and writing activities to promote independent communication and comprehension, as well as into literacy lessons by assigning reading exercises where students dictate summaries and receive AI feedback.

\subsection{Adaptive Learning Systems}

Adaptive learning systems, powered by artificial intelligence and machine learning, are revolutionizing education by providing personalized learning experiences tailored to individual student needs. These systems continuously assess student performance and adjust the difficulty and focus of learning materials, making them particularly effective for students with special needs by offering personalized support and addressing diverse learning styles \cite{AkashJagtap2025, LavanyaME2024}. Platforms like Carnegie Learning and DreamBox exemplify the application of adaptive learning in math and reading, while Notebook uses machine learning to deliver personalized learning experiences \cite{AkashJagtap2025, SiqiLi2025}. The integration of AI in education allows for real-time feedback, dynamic content adjustment, and predictive analytics, which enhance student engagement, retention, and academic performance \cite{LavanyaME2024, NurHakim2024}. Studies have shown that adaptive learning algorithms significantly improve student outcomes, with higher engagement and satisfaction rates compared to traditional learning models \cite{NurHakim2024, GalinaYulina2024}. These systems utilize data-driven insights to create a more inclusive and accessible learning environment, catering to the diverse needs of modern students \cite{AkashJagtap2025, TRaghavendraGupta2024}. However, challenges such as the complexity of technology development, digital infrastructure gaps, and the need for teacher training remain \cite{AkbarMaulana2025, EmmanuelIdowu2024}. Despite these challenges, the potential of adaptive learning systems to democratize education and foster lifelong learning is immense, as they allow students to progress based on mastery rather than age or grade level \cite{EmmanuelDumbuya2024, EmmanuelIdowu2024}. As educational institutions continue to explore these technologies, the synergy of personalized and adaptive learning approaches promises to transform education by meeting learners where they are and supporting their individual learning journeys \cite{GalinaYulina2024, EmmanuelIdowu2024}.

\paragraph{Summary and Classroom Implications.}
Adaptive systems provide continuous performance monitoring and customized content delivery. Educators can integrate these tools into core lessons, allowing diverse learners to progress at their own pace while maintaining alignment with curriculum goals.

\subsection{Virtual and Augmented Reality (VR/AR)}

Virtual and Augmented Reality (VR/AR) technologies have emerged as transformative tools in education, particularly for students with special needs, by creating immersive learning environments that simulate real-world scenarios. These technologies are especially beneficial for students with Autism Spectrum Disorder (ASD), as they provide a safe and controlled environment for practicing social interactions and navigating complex environments. VR interventions have been shown to enhance social skills in autistic children by offering engaging and adaptable experiences that traditional therapies often fail to provide \cite{YueminZhu2024, ChiaraFailla2024}. The immersive nature of VR allows for the creation of realistic simulations that can improve social, cognitive, and linguistic abilities, making it a promising tool for therapeutic interventions in ASD \cite{ChiaraFailla2024}. Additionally, AR tools like the Augmenta11y app cater to dyslexic students by displaying content in accessible formats, thus addressing diverse learning needs and making education more inclusive \cite{DLinettSophia2024}. Examples of AI-integrated tools such as Google Glass and Lexplore further assist students with reading difficulties and social challenges, demonstrating the potential of these technologies to support personalized learning experiences \cite{ChiaraFailla2024}. Despite the high costs and technical barriers associated with VR/AR, their ability to create dynamic, learner-centric environments underscores their transformative potential in education \cite{DLinettSophia2024, SanchitVashisht2024}. Future research should focus on making these technologies more affordable and accessible, as well as exploring their long-term educational impacts \cite{DLinettSophia2024, SanchitVashisht2024}. The integration of AI with VR/AR can further enhance these experiences by allowing real-time adaptation to individual needs, thereby promoting a more supportive and less anxiety-inducing learning environment for students with special needs \cite{ChiaraFailla2024}. Overall, VR and AR are poised to become integral tools in modern education, offering innovative solutions to the challenges faced by students with special needs and paving the way for more inclusive and effective educational practices \cite{JibranZeb2024, ErenDanyelUar2024}.

\paragraph{Summary and Classroom Implications.}
VR/AR tools create immersive experiences tailored to students with developmental or sensory processing needs. Teachers can deploy these technologies during social skills training or personalized literacy tasks to enhance emotional and cognitive engagement. Immersive language simulations can be used to practice real-life interactions, such as virtual shopping or greetings, allowing structured lesson alignment with specific learning outcomes.

\subsection{Facial Expression and Emotion Recognition}

AI-powered facial recognition technologies are increasingly being utilized to identify and interpret emotional cues, offering significant benefits in educational settings, particularly for students with Autism Spectrum Disorder (ASD). These technologies help address the unique social and communication challenges faced by students with ASD by enabling educators to better understand and respond to their emotional states. AI-enhanced tools that integrate Internet of Things (IoT) capabilities and emotional intelligence algorithms, such as those using Haar-cascade Python libraries and Convolutional Neural Networks (CNN), have been developed to capture real-time facial expressions and monitor stress levels through Galvanic Skin Response (GSR) sensors \cite{DishoreShunmughamVanaja2025, SureshKallam2024}. This allows for timely interventions and personalized support, enhancing social engagement and interaction skills in children with ASD. Additionally, Human-Robot-Game platforms have been refined to improve facial recognition metrics, providing accurate measurements of attention and emotional responses, which are crucial for assessing the socio-educational progress of children with ASD \cite{NayethIdalidSolrzanoAlcvar2024}. The use of AI in education extends beyond ASD, as emotion recognition systems employing deep learning techniques can classify student emotions into basic categories, offering insights that help create supportive learning environments \cite{DebajyotiChatterjee2024}. Transfer learning models like VGG-16 and ResNet-18 have shown promise in classifying student emotions, although challenges such as overfitting and the need for improved generalization remain \cite{TitaHerradura2024}. Despite the potential of these technologies, issues such as data privacy, ethical considerations, and the need for culturally sensitive systems are critical challenges that need to be addressed to ensure the effective implementation of AI in educational settings \cite{VaibhavJindal2025, CliaLlurba2024}. Overall, while AI-driven facial recognition technologies hold promise for enhancing educational outcomes, particularly for students with ASD, ongoing research and development are necessary to optimize their effectiveness and address existing limitations \cite{SofiaKotsi2025}.

\paragraph{Summary and Classroom Implications.}
Emotion recognition technologies can support classroom emotional climate monitoring. When ethically and appropriately applied, they allow teachers to better understand students' unspoken needs and respond with timely interventions.

\subsection{Interactive Robots}

Interactive robots are increasingly being integrated into educational settings to support students with physical or cognitive disabilities, offering both educational and emotional assistance. These robots, equipped with AI-driven features like speech recognition, gesture analysis, and personalized feedback mechanisms, are designed to create engaging and accessible learning experiences for students with neurodevelopmental disorders \cite{AndyIsmail2024, AndreaRHarkinsBrown2025, ParthenopiPapalexandratou2024}. For instance, the adaptive AI robot developed at Universitas Darwan Ali significantly improved academic comprehension by 30\% and achieved an 82\% satisfaction rate among students with learning disabilities, demonstrating its effectiveness in inclusive education \cite{AndyIsmail2024}. Similarly, the \textit{Sahayak} app utilizes augmented reality and AI to tailor learning experiences to the specific needs of children with intellectual disabilities, enhancing their reading and comprehension skills through interactive elements \cite{TejasGirhe2024}. Moreover, educational robots like those in the Emorobot Project are used to foster social skills in children with autism spectrum disorders by recognizing emotions and facilitating peer connections, thereby reducing social isolation and promoting interpersonal skills \cite{FrancescoLoSchiavo2024}. The transformative role of AI and robotics in special education is further highlighted by their ability to customize educational content to individual learning styles, thus enhancing strategy efficacy through data-driven insights \cite{PrincyPappachan2024}. Despite these advancements, ethical considerations, potential biases, and privacy concerns remain critical issues that need to be addressed to ensure the responsible integration of AI in special education \cite{AndreaRHarkinsBrown2025, AyeAlkan2024}. Overall, interactive robots represent a promising avenue for creating inclusive educational environments that cater to the diverse needs of students with disabilities, fostering both academic success and emotional well-being \cite{MasoodaModak2024, DrVSheejaVayola2023}.

\paragraph{Summary and Classroom Implications.}
Interactive robots can function as social and instructional partners for neurodivergent students. Educators can use them during targeted interventions, combining human interaction with AI-based feedback to foster motivation and peer collaboration.

\subsection{Early Intervention and Diagnostic Tools}

AI-powered diagnostic tools have significantly advanced the early identification and intervention of learning disabilities and developmental delays, offering promising solutions for educational and healthcare settings. These tools utilize machine learning algorithms to analyze student performance and behavioral data, effectively detecting challenges such as dyslexia, dyscalculia, ADHD, and autism spectrum disorder (ASD) \cite{DrShilpiJain2024, ManaraAlHamieli2024, SSanthiya2023}. For instance, AI-driven platforms can identify struggling students and provide adaptive resources tailored to their specific needs, thereby facilitating timely and targeted support strategies \cite{HasanPhudinawala2024, DelagrammatikaGaryfallia2024}. The integration of multi-sensory technologies in diagnostic systems further enhances the assessment of cognitive, motor, and social skills, providing personalized interventions that improve developmental outcomes \cite{LohanKavindaBandara2024}. The use of AI in early diagnosis is particularly beneficial for conditions like dyslexia, where early intervention can mitigate the disorder's impact on educational, psychological, and social dimensions \cite{SafaAbduljalilAwadhBaitSaleem2025}. Moreover, AI applications in mobile platforms have shown to expedite the diagnosis and classification of ASD, improving accuracy and accessibility, especially in resource-limited areas \cite{ManaraAlHamieli2024}. Despite the potential of AI tools, their implementation requires careful consideration of ethical issues, such as data privacy and potential biases, to ensure fair and inclusive educational practices \cite{RosdianaRosdiana2024}. Overall, the deployment of AI in early detection and intervention not only enhances diagnostic accuracy but also empowers educators and healthcare professionals to provide customized support, ultimately fostering better educational and developmental trajectories for children with learning disabilities \cite{NKalyani2024, AndreaBurgess2024}.

\paragraph{Summary and Classroom Implications.}
Early diagnostic AI systems assist in identifying learning challenges promptly. Educators can use their outputs to craft individualized education plans and collaborate more effectively with caregivers and specialists.

\subsection{Blended Learning Models}

Blended learning models, which integrate traditional teaching methods with AI-driven technologies, are increasingly recognized for their ability to create inclusive and flexible educational environments. These models are particularly effective in inclusive classrooms, where they cater to diverse learning needs by enhancing accessibility and engagement, ensuring that all students can participate meaningfully. The integration of AI with Universal Design for Learning (UDL) frameworks is instrumental in maximizing learning outcomes for students with disabilities, as AI-driven adaptive learning systems significantly outperform traditional methods in personalizing educational experiences for disabled learners \cite{SMahalakshmi2025}. The AI-enabled hybrid learning models facilitate interactions in both face-to-face and online settings, allowing students to benefit from network resources for independent learning, thereby improving learning effects \cite{WenzhengCai2024}. Moreover, the use of AI in blended learning environments, such as through gamification and personalized learning paths, has shown significant improvements in educational outcomes, as evidenced by a study in Shanxi, China, where students experienced a 25\% higher improvement in vocabulary learning and a 30\% increase in reading comprehension \cite{ZiheWu2024}. The integration of AI and multiple intelligences frameworks further enhances personalized instruction, promoting diverse talents and augmenting creativity and problem-solving capabilities \cite{SureshPalarimath2024}. Despite these advancements, challenges such as equity and access persist, particularly for underprivileged students who may face barriers due to lack of internet connectivity and digital resources \cite{RobertMulenga2024}. However, the potential of AI and blended learning to transform educational practices by offering personalized, engaging, and adaptable learning experiences remains significant, pointing towards a future where education is more accessible and inclusive for all students \cite{RazaASiddiqui2024, RobertMulenga2024, NajibaMorrar2024}.

\paragraph{Summary and Classroom Implications.}
AI-enabled blended learning supports differentiated instruction by balancing tech-mediated and teacher-led activities. This hybrid model lets teachers adapt pace and format, improving engagement and outcomes across diverse learning profiles.

\subsection{AI-Driven Accessibility Features}

AI-driven accessibility features are revolutionizing educational settings by enhancing independence and inclusivity for students with disabilities. These technologies, such as real-time translation, image recognition, and personalized interfaces, are pivotal in creating adaptive learning environments that cater to individual needs. AI-powered braille devices, for instance, enable visually impaired students to learn independently, showcasing the transformative potential of AI in education \cite{MunikrishnaiahSundaraRamaiah2024, NnaemekaValentineEziamaka2024}. Automated captioning systems and text-to-speech converters are crucial for students with hearing or visual impairments, ensuring seamless access to educational content \cite{ErenKamber2025}. AI-driven speech recognition and emotion recognition systems have demonstrated significant improvements in communication efficiency and user satisfaction, further supporting students with disabilities in educational contexts \cite{GoenawanBrotosaputro2024}. Moreover, AI technologies like object detection, navigation, and optical character recognition (OCR) provide comprehensive support for visually impaired individuals, facilitating safe navigation and access to written information \cite{ANRamyaShree2024}. The integration of AI in educational tools also includes adaptive learning systems and generative AI chatbots, which create personalized learning paths and support real-time communication, thus transforming the learning process for students with disabilities \cite{XhulioMitre2024}. Additionally, multisensory AI systems enhance accessibility by integrating sensory modalities such as vision, hearing, and touch, offering adaptive solutions that bridge accessibility gaps in education \cite{MuraliKrishnaPasupuleti2025}. Despite these advancements, challenges such as data privacy, algorithmic bias, and the need for inclusive design practices remain critical considerations for the effective deployment of AI-driven accessibility technologies \cite{NnaemekaValentineEziamaka2024}. Overall, AI's role in educational accessibility is not only about technological innovation but also about fostering equity and inclusion, ensuring that all learners can participate meaningfully in educational activities \cite{MBhakiyasri2024, AllieZombron2024}.

\paragraph{Summary and Classroom Implications.}
Accessibility features like Braille displays, automated captions, and OCR allow students with disabilities to access content independently. Educators should ensure these tools are integrated seamlessly into class routines, ensuring equitable participation in all academic tasks.

\paragraph{Section Summary:} Section 4 provided a deep dive into AI-driven assistive technologies, showcasing their potential to personalize learning and support a diverse range of special needs. The summaries at the end of each tool type offer immediate links to classroom practice, reinforcing the inclusive aims of this chapter.

\section{Influence of AI on Early Educators Roles and Workloads}
\subsection{AIs Role in Reducing Workload}

The integration of AI tools in early education settings is transforming the traditional roles of teachers by alleviating some of their workload and enhancing educational practices. AI applications, such as Learning Management Systems (LMS) and educational software, are increasingly used to automate routine tasks like grading and data entry, allowing teachers to focus more on personalized instruction and student engagement \cite{CaseStudyofUsingAIToolsinPrimarySchool2025, TheRoleofAIinEducation2025}. These tools not only save time but also provide valuable insights into student performance, helping educators identify areas where students struggle and tailor their teaching strategies accordingly \cite{AbdulQayyum2024}. However, while AI offers significant benefits, it also presents challenges, such as the inability to interpret social cues and the need for human interaction, particularly in early childhood education \cite{AbdulQayyum2024}. Teachers generally perceive AI tools positively, recognizing their potential to enhance learning outcomes and support personalized feedback, although concerns about technical challenges and the need for adequate training persist \cite{WeipengYang2024}. Moreover, AI's role in fostering computational thinking (CT) and problem-solving skills is increasingly emphasized, with educational programs integrating AI literacy to prepare young learners for a tech-driven future \cite{WeipengYang2024}.

AI is also increasingly being integrated into special education to streamline documentation and reduce the administrative burden on educators, thereby allowing them to focus more on student interaction. AI technologies, such as natural language processing and speech-to-text tools, are being utilized to draft reports, summarize data, and suggest educational goals, which significantly cuts down the bureaucratic workload associated with tasks like Individualized Education Program (IEP) reports and behavior logs \cite{AndreaRHarkinsBrown2025, SamanthaRGoldman2024}. Surveys indicate that most special educators are optimistic about AI's potential to alleviate administrative tasks, thus enabling more direct engagement with students \cite{SamanthaRGoldman2024}. Early pilot programs have demonstrated that AI-powered lesson-planning tools can save teachers several hours each week by suggesting differentiated activities and generating adaptable lesson drafts \cite{MunikrishnaiahSundaraRamaiah2024, WasimAhmad2024}. These tools not only enhance teaching efficacy but also support personalized learning experiences tailored to the unique needs of students with disabilities \cite{AyeAlkan2024, LynnMScott2024}. AI's role extends beyond administrative efficiency; it also fosters inclusivity and accessibility by providing customized, responsive support that addresses diverse learning needs \cite{MunikrishnaiahSundaraRamaiah2024, SilvioMarcelloPagliara2024}. However, the integration of AI in education must be approached with caution, considering ethical concerns such as privacy, bias, and the digital divide \cite{SilvioMarcelloPagliara2024, RituArya2024}. Despite these challenges, AI's potential to revolutionize special education is significant, offering scalable solutions that enhance both educational outcomes and teacher productivity \cite{WasimAhmad2024, RumyanaPapancheva2024}. As AI continues to evolve, its thoughtful implementation, guided by empirical research and professional oversight, is crucial to ensuring that it truly benefits students with disabilities and supports educators in their roles \cite{AndreaRHarkinsBrown2025, DonDMcMahon2024}.

Table \ref{tab:aifeatures} summarizes the AI features and their impact on teaching roles.

\begin{table}[h]
    \centering
    \caption{AI Features and Their Impact on Teaching Roles}
    \begin{tabular}{p{4cm} p{7cm} p{2cm}}
        \toprule
        \textbf{AI Feature} & \textbf{Impact on Teaching Roles} & \textbf{References} \\
        \midrule
        Adaptive Learning Platforms & Personalizes learning experiences, reduces administrative tasks, and enhances student engagement & \cite{AndreaRHarkinsBrown2025, SilvanaMariaAparecidaVianaSantos2024, SamanthaRGoldman2024} \\
        Virtual Assistants & Facilitates communication and social interaction for students with special needs & \cite{AlexandraGuedes2024, MehdiRostami2024} \\
        Language Translation Tools & Breaks down language barriers in multilingual classrooms & \cite{ElizabethAngoFomusoEkellem2024} \\
        Automated Grading and Lesson Planning & Reduces time spent on administrative tasks, allowing focus on instructional activities & \cite{EdithNamutebi2024, SamanthaRGoldman2024, MercedesSantos2024} \\
        AI-Powered Teaching Assistants & Provides real-time feedback and supports personalized learning experiences & \cite{FangzhouJin2024, nez2024} \\
        Ethical Considerations & Ensures AI tools are developed and implemented in an ethical and inclusive manner & \cite{MuhammadAsyari2024, SilvioMarcelloPagliara2024} \\
        \bottomrule
    \end{tabular}
    
    \label{tab:aifeatures}
\end{table}

\subsection{Shifting Toward Personalized Teaching}
The role of teachers is evolving from traditional content delivery to a more personalized and facilitative approach, where they act as guides or coaches for students' individual learning journeys. AI technologies, such as intelligent tutoring systems and adaptive learning platforms, enable personalized learning experiences by analyzing student data and tailoring educational content to individual needs \cite{WaiYieLeong2025, CindyFitriLaksono2024}. This allows teachers to focus on more complex instructional duties, such as designing creative learning experiences and providing one-on-one attention to students who need extra help or enrichment \cite{RegitaAmaliaSeptiani2025, TarunKumarVashishth2025}. The integration of AI in education also automates administrative tasks, such as grading and attendance tracking, freeing up teachers to concentrate on fostering a supportive and engaging learning environment \cite{AbdulKodir2025, CindyFitriLaksono2024}. Teachers can orchestrate classrooms into centers, where AI-powered apps engage students at their own pace, while the teacher works intensively with small groups on targeted skills, effectively acting as an "inclusion orchestrator" \cite{ErnaWidyasari2024}. Despite the transformative potential of AI, teachers remain irreplaceable in building emotional connections, instilling moral values, and guiding students' character development \cite{RegitaAmaliaSeptiani2025, AzizMimoudi2024}. The symbiotic relationship between AI tools and educators is crucial for creating inclusive and effective education, emphasizing the need for professional development and policy reforms to maximize the benefits of AI integration \cite{TarunKumarVashishth2025, ASSasipriya2024}. However, challenges such as inequitable access to technology, potential biases in algorithms, and data privacy issues must be addressed to ensure ethical and equitable AI implementation in education \cite{ErnaWidyasari2024, ASSasipriya2024}.

The integration of AI in educational assessment and feedback is also transforming traditional teaching practices by providing instant, personalized feedback, which empowers students to take greater ownership of their learning and progress at their own pace. This shift allows teachers to focus on higher-order feedback and more complex educational tasks, such as addressing subtle misunderstandings and emotional barriers to learning, which AI cannot detect \cite{JamesHutson2024, SmithaShivshankar2024}. AI tools, such as generative AI and large language models, are being used to automate routine grading tasks and generate personalized feedback, freeing educators to engage in more meaningful interactions with students, such as mentoring and facilitating hands-on learning activities \cite{JamesHutson2024, MichaelCooperStachowsky2024}. This approach not only enhances student engagement but also improves learning outcomes by allowing teachers to act as facilitators or learning designers, tailoring the educational environment to better meet individual student needs \cite{JamesHutson2024, SElizaFemiSherley2024}. However, the adoption of AI in assessment also presents challenges, such as concerns over academic integrity, algorithmic bias, and equitable access, which educators must navigate to ensure ethical and effective use \cite{WIllMiller2024, KudakwasheManokore2024}. Despite these challenges, AI's ability to provide immediate, consistent feedback and analyze vast amounts of data can lead to more sophisticated and adaptive learning environments, supporting dynamic and personalized educational experiences \cite{RobertasDamaeviius2024, MawuliApetorgbor2024}. The AI Assessment Scale (AIAS) and other frameworks have been developed to integrate AI into educational assessments effectively, promoting academic integrity while leveraging AI's potential to enhance learning experiences \cite{LeonFurze2024}. Overall, AI's role in education is not to replace teachers but to complement their expertise, allowing them to focus on the "brain work" and "heart work" of teaching, thereby fostering deeper understanding and critical thinking skills among students \cite{RobertasDamaeviius2024, MohdRushidiMohdAmin2024}.

\subsection{Challenges in Implementation}

While the potential for AI to reduce workload and transform teaching roles is attractive, implementing these changes in real classrooms comes with significant challenges. One major challenge is teacher training and confidence. Many educators, despite being familiar with AI tools, lack formal training in effectively utilizing these technologies, leading to underutilization or misapplication in classroom settings \cite{JulieADelello2025}. The rapid development of AI technologies, such as Generative AI, necessitates that teacher educators and pre-service teachers acquire AI literacies, including understanding AI as a teaching tool and interpreting AI outputs for pedagogical decisions \cite{ChristopherNeilPrilop2024}. However, professional development programs specific to AI in education are still emerging and not yet widespread, which contributes to teachers' anxiety and reluctance to fully embrace AI tools \cite{JulieADelello2025, ChristopherNeilPrilop2024}. Studies have shown that when teachers receive coaching on integrating AI into their practice, their stress levels decrease, and they are more likely to use AI in ways that enhance teaching efficiency and save time \cite{JulieADelello2025, HeniMulyani2025}. Moreover, the evolving role of teachers in AI-enhanced classrooms underscores the need for professional development that emphasizes human-centric skills such as empathy and critical thinking, alongside technical proficiency \cite{TarunKumarVashishth2025}. The lack of comprehensive institutional policies and ethical frameworks further complicates the integration process, as teachers often have to navigate these challenges independently \cite{JulieADelello2025, NanXiao2025}. To address these issues, it is crucial to develop training programs that not only focus on the operational aspects of AI tools but also on interpreting AI analytics and making informed pedagogical decisions \cite{WeiweiZhang2024}. Such initiatives would help mitigate the risks of either mistrusting or over-trusting AI recommendations, ensuring that the benefits of AI in education are fully realized \cite{YPan2024}.

Another challenge is infrastructure and support. Many educational institutions lack the necessary hardware, software, and IT support to seamlessly incorporate AI systems, leading to increased stress for teachers who must deal with unreliable internet connections and outdated devices. This issue is compounded by the frequent crashes of AI systems, such as grading software, and the insufficient number of devices for students, which forces teachers to spend valuable time troubleshooting and rearranging lessons, thereby increasing their workload rather than alleviating it \cite{AbdulKodir2025, WaiYieLeong2024}. Financial constraints, especially in rural or underserved areas, limit schools' ability to afford comprehensive AI solutions, as many AI tools require costly licenses or subscriptions. This financial barrier exacerbates educational inequities, as well-resourced schools can access cutting-edge AI tools while others cannot, deepening the digital divide \cite{WaiYieLeong2024, PravinTambat2024}. The lack of infrastructure is a recurring theme across various educational contexts, from engineering education to secondary education in India, where strategic investment and policy support are necessary to overcome these barriers \cite{Zhang2024, PravinTambat2024}. Additionally, ethical concerns surrounding AI, such as data privacy and algorithmic bias, further complicate its implementation, necessitating the development of ethical frameworks and teacher training to ensure responsible use \cite{NanXiao2025, KonstantinosKotsis2024}. To address these challenges, targeted technical solutions and strategic investments in infrastructure and training are essential to create an inclusive and sustainable learning ecosystem that leverages AI's potential while ensuring equitable access for all students \cite{WaiYieLeong2024, OluwaseyiAinaGboladeOpesemowo2024}.

There are also time and curricular constraints. Teachers often face increased workloads during the initial phase of AI adoption, as they must adapt lesson plans and familiarize themselves with new systems, which can be particularly burdensome in environments with limited technological infrastructure and training, such as rural areas \cite{KamarullahKamarullah2024}. The restructuring of class time and space to accommodate AI-driven methods, like personalized software centers, can be challenging under rigid curriculum schedules or large class sizes, potentially leading to the abandonment of AI tools if they are perceived as interfering with curriculum coverage \cite{MercedesSantos2024}. Despite these challenges, AI tools offer significant benefits, such as enhancing lesson planning and teaching effectiveness through generative AI, which can alleviate time constraints and improve the quality of teaching \cite{FrankKehoe2023, HeniMulyani2025}. However, successful integration requires careful management of technical limitations and the provision of adequate teacher training to ensure that AI complements rather than complicates traditional teaching methods \cite{ChenxinZhang2024, ChenghaoWang2024}. Teachers' perceptions of AI's usefulness and ease of use are crucial for its acceptance, and professional development initiatives are necessary to address gaps in technological content knowledge and experience \cite{IrisHeungYueYim2024}. Moreover, AI's potential to personalize learning and automate administrative tasks can free up time for more teaching-focused activities, although this necessitates new training and technological adaptations \cite{MercedesSantos2024}. To ensure equitable and responsible AI integration, it is essential to address ethical concerns, such as data privacy and algorithmic bias, and to develop inclusive policies that support diverse educational contexts \cite{ExploringAIDrivenPedagogicalTools2024, MercedesSantos2024}.

Finally, cultural and ethical challenges can affect implementation. One primary concern is the skepticism among teachers and parents regarding AI's potential to diminish the teacher's role and misjudge students, which stems from fears of job replacement and data privacy issues \cite{MercedesSantos2024, DrSeemaYadav2024}. The opacity of AI decision-making processes and algorithmic bias further exacerbate these concerns, as they can perpetuate inequalities and affect assessment outcomes \cite{OkanBulut2024, AmineAmmar2025}. Teachers often feel their professional judgment is undermined by AI systems, which can recommend student groupings or identify weaknesses, leading to a perceived loss of control \cite{MercedesSantos2024}. To address these issues, it is crucial that AI tools are explainable and that teachers are involved in their development or selection, ensuring transparency and building trust among stakeholders \cite{OkanBulut2024, DrSeemaYadav2024}. Additionally, the ethical use of AI in education requires maintaining integrity and equity, as well as safeguarding privacy and data security \cite{AmirAhmadDar2024, DrSeemaYadav2024}. The cultural responsibility of AI systems is also paramount, necessitating their adaptability to multicultural societies and the implementation of regulatory measures to promote cultural sensitivity \cite{NataliaOzegalskaLukasik2023}. Furthermore, the ethical challenges associated with large-scale data labeling, such as bias and privacy violations, highlight the need for transparency and accountability in AI data management \cite{AmineAmmar2025}. Addressing these challenges involves collaboration among educators, policymakers, and stakeholders to ensure AI's responsible and effective use in education, ultimately promoting equitable, inclusive, and effective teaching and learning practices \cite{DrSeemaYadav2024, AltieresdeOliveiraSilva2024}.

\subsection{Ethics \& TeacherAI Dynamics}
The infusion of AI into the classroom also brings ethical considerations \cite{Fitas2025AI} and a redefinition of teacherAI dynamics that educators must actively manage. One ethical priority is maintaining student privacy and dignity. Early childhood and primary students cannot consent or advocate for their data rights. The responsibility falls, therefore, on teachers and schools to ensure compliance with privacy laws such as FERPA and GDPR, safeguarding sensitive student information like performance data, behavior logs, and recordings \cite{PragyaMishara2024, AdemYILMAZ2024, PElantheraiyan2024}. The ethical implications of AI in education extend beyond privacy to include concerns about bias and accountability, necessitating proactive policies that prioritize ethical standards to prevent exacerbating existing inequalities \cite{PragyaMishara2024}. In Indonesia, for example, the implementation of AI in education highlights the need for transparency in AI algorithms and decision-making processes, as well as the importance of maintaining human control and ensuring inclusivity for all students, including those with special needs \cite{MusidiansyahOttoSyaidina2024}. The integration of AI in primary education, while promising enhanced learning experiences, raises ethical questions about child privacy, as AI-driven platforms collect sensitive data from personal identifiers to learning preferences \cite{PElantheraiyan2024}. To address these concerns, a comprehensive framework that includes privacy-preserving protocols, bias mitigation strategies, and inclusivity measures is essential \cite{YousefJaberJamelAlawneh2024}. Additionally, the need for user-centric privacy controls and tailored transparency strategies is underscored by stakeholder analyses, which reveal the importance of education and awareness in influencing data ownership and risk assessment \cite{MollyCampbell2025}. The development of ethical AI systems in education should be guided by principles that ensure safe, transparent, and ethical use, aligning with institutional strategies and minimizing potential risks such as privacy violations \cite{FranciscoJosGarcaPealvo2024}. A notable case in Milan involving AI-based e-proctoring during online exams illustrates the ethical complexities of educational AI systems \cite{bai2024re}. The system collected biometric data and used automated algorithms to flag suspicious behavior, ultimately drawing legal scrutiny under the GDPR for violations related to consent, transparency, and data minimization. This case underscores the critical need for educational AI toolsespecially those supporting special needs or real-time translationto be implemented with strong safeguards. AI applications that assist students with disabilities or language barriers must be designed with the same level of privacy consideration, ensuring compliance with ethical standards and legal frameworks. Developers, educators, and institutions should co-create these tools through multi-stakeholder engagement, balancing innovation with respect for student rights and agency.

Another ethical concern is algorithmic bias and fairness. Bias in AI arises from skewed training data and societal influences, which can lead to systematic errors in decision-making processes, affecting marginalized communities disproportionately \cite{TVenkatNarayanaRao2025, AgariadneDwinggoSamala2024}. In educational settings, AI tools like reading tutors and predictive analytics can perpetuate inequalities if they are not designed with diverse datasets, potentially misjudging the progress of English learners or students with disabilities \cite{OkanBulut2024, AlpDulundu2024}. For instance, an AI reading tutor trained mostly on English speakers without disabilities might not accurately gauge the progress of an English learner or a student with a speech impairment, potentially giving misleading feedback. This issue is compounded by the opacity of AI decision-making, which can obscure the biases embedded within these systems, leading to unfair assessment outcomes and stigmatization of students labeled as "high risk" based on historical biases \cite{OkanBulut2024, PragyaMishara2024}. To address these challenges, it is crucial for educators and developers to implement fairness-aware learning and algorithmic transparency, ensuring that AI systems are both equitable and effective \cite{AgariadneDwinggoSamala2024, KhalidaWalidNathim2024}. Moreover, diverse representation in AI development and community involvement are essential to mitigate biases and enhance cultural sensitivity, thereby fostering trust and ensuring equitable access to technology \cite{NoticePasipamire2024, AdamLockwood2024}.

The integration of AI in education is transforming traditional pedagogical practices, yet it underscores the necessity of maintaining the teacher's role as the ultimate decision-maker. AI tools offer significant benefits, such as personalized learning, administrative efficiency, and enhanced classroom management, which can improve student engagement and learning outcomes \cite{RegitaAmaliaSeptiani2025, RoohUllah2024, SuchetaYambal2024}. However, the role of teachers remains irreplaceable, particularly in building emotional connections, instilling moral values, and guiding character development, which AI cannot replicate \cite{RegitaAmaliaSeptiani2025}. The U.S. Department of Education and other educational authorities advocate for a "human-in-the-loop" approach, where AI serves as an advisory tool, and educators vet AI recommendations to ensure ethical and contextual considerations are applied \cite{JulieADelello2025, ChenghaoWang2024}. This approach is crucial, as AI systems may overlook nuances such as personal student circumstances, which teachers can address through their holistic understanding \cite{RegitaAmaliaSeptiani2025, AlpDulundu2024}. Despite AI's potential to reduce stress through automation, concerns about increased anxiety, social isolation, and ethical implications persist, highlighting the need for professional development in AI literacy and ethical considerations \cite{JulieADelello2025, MehwishRaza2024}. Moreover, the rapid adoption of AI tools necessitates retraining and upskilling teachers to effectively integrate AI with traditional teaching methods, ensuring that AI serves as a supplement rather than a replacement \cite{rus12024, MehwishRaza2024}. The balance between leveraging AI's capabilities and maintaining human oversight is essential to fostering an inclusive and effective educational environment that aligns with ethical standards and promotes student well-being \cite{AlpDulundu2024, YTalgatov2024}.

The changing teacherAI dynamic also raises the question of professional identity: teachers may wonder if AI will replace aspects of their role or even render some positions obsolete. AI excels in data processing and personalized learning, yet lacks the human qualities of empathy, adaptability, and moral judgment that are essential in education \cite{RegitaAmaliaSeptiani2025, TarunKumarVashishth2025}. Teachers are seen as irreplaceable in building emotional connections, instilling moral values, and guiding character development, roles that AI cannot fulfill \cite{RegitaAmaliaSeptiani2025}. The concept of "AI as the new TA" suggests that AI can amplify teachers' effectiveness by handling administrative tasks and providing personalized learning experiences, allowing teachers to focus on fostering a supportive and engaging learning environment \cite{TarunKumarVashishth2025}. However, the successful integration of AI requires teachers to be actively involved in the selection and customization of AI systems, which fosters trust and effective use \cite{EseEmmanuelUwosomah2025}. Involving teachers in AI integration can transform AI from a potential adversary into a collaborative colleague, enabling teachers to use AI for brainstorming and managing different student groups, akin to team-teaching \cite{EseEmmanuelUwosomah2025, HyeonjeongLEE2024}. Conversely, if AI is introduced primarily as a cost-cutting measure, it may lead to resistance and an adversarial dynamic \cite{RegitaAmaliaSeptiani2025}. The evolving role of teachers in AI-enhanced environments positions them as facilitators of learning, emphasizing the need for professional development and policy reforms to maximize AI's benefits while maintaining human-centric skills like empathy and creativity \cite{TarunKumarVashishth2025}. Teacher training is crucial to enhance AI-related skills and promote human-computer collaboration models, ensuring that AI integration aligns with educational and societal values \cite{RuiWu2024, ChristopherNeilPrilop2024}. As AI becomes more integrated into education, ongoing discussions are necessary to reflect on and fulfill the evolving roles of teachers, ensuring that their professional identity remains central amidst technological advancements \cite{HyeonjeongLEE2024, BenjaminDParker2024}.

Finally, theres also the duty to prepare students for a world with AI  which means teachers themselves model a critical and balanced approach to technology. In preparing students for a world increasingly influenced by AI, educators must model a critical and balanced approach to technology, as highlighted across various studies. Teachers play a pivotal role in demystifying AI by engaging students in discussions about how AI works, its applications, and its limitations, thereby fostering digital literacy and ethical awareness. For instance, a teacher might explain that while a tablet game can aid in reading, it may also make errors, encouraging students to critically evaluate AI outputs and maintain agency in their learning process\cite{rus12024, YPan2024}. This approach aligns with the need for educators to develop AI literacy, enabling them to integrate AI tools effectively and responsibly into their teaching practices\cite{PrasartNuangchalerm2024}. The integration of AI in education offers transformative potential, such as personalized learning experiences and improved student engagement, as evidenced by increased test scores and participation rates in AI-enhanced classrooms\cite{SilvanaSilva2025}. However, ethical concerns, such as data privacy, algorithmic bias, and the potential disruption of traditional student-teacher dynamics, must be addressed to ensure AI's responsible use in education\cite{JaredMomanyiMauti2025, YPan2024}. Moreover, the disparity in AI adoption between resource-rich and under-resourced communities highlights the need for equitable access to AI technologies to bridge educational gaps\cite{OluwaseyiAinaGboladeOpesemowo2024}. Teacher training programs must incorporate reflective practices and situated learning to prepare educators for these challenges, promoting a human-centered approach to AI integration\cite{FranciscoSereoAhumada2024}. As AI becomes an indispensable educational tool, ongoing professional development and interdisciplinary collaboration are crucial for empowering educators to navigate the evolving landscape of AI in education, ultimately preparing students for a technology-driven future\cite{SeemaYadav2024}.

\paragraph{Section Summary:} Section 5 emphasized that while AI can assist in reducing teacher burden and supporting individualized learning, the educators role as a guide, decision-maker, and emotional support remains essential. Strategic training, ethical frameworks, and infrastructure investment are key to ensuring AI adoption enhances rather than replaces teaching roles.

\section{General Outcomes}

Integrating AI into early inclusive education has demonstrated numerous positive outcomes, particularly in enhancing student engagement and learning experiences. AI tools, such as interactive tutors and gamified learning applications, are designed to be adaptive and responsive, which significantly captivates young learners' attention compared to traditional static materials. This adaptability is crucial for maintaining student interest and preventing frustration, especially for students with special needs, such as those with attention deficit hyperactivity disorder (ADHD), who benefit from AI tutors that provide instant feedback and rewards for small successes, thereby sustaining motivation throughout tasks \cite{SriSukasih2024, AndyIsmail2024}. In multilingual settings, AI's translation and language support capabilities enable English learners and other language-minority students to engage with content more effectively, allowing them to participate actively in class and interact confidently with peers, thus improving overall class participation rates \cite{MansiGupta2024}. Furthermore, AI's role in inclusive education extends to supporting students with specific learning disorders, such as dyslexia and dyscalculia, by offering personalized learning experiences that adjust to individual abilities and learning styles, thereby enhancing literacy skills and conceptual understanding \cite{SriSukasih2024, SYARIFMAULIDIN2024}. Despite these benefits, challenges remain, such as ensuring content accuracy, age appropriateness, and cultural relevance, as well as addressing technological infrastructure disparities, particularly in regions like Ukraine and India, where educational systems are still evolving to integrate AI effectively \cite{rus22024, MansiGupta2024}.

AI-driven adaptive learning systems have demonstrated significant potential in enhancing foundational skills such as early literacy and numeracy by tailoring educational experiences to individual student needs. These systems utilize advanced machine learning algorithms and real-time feedback mechanisms to provide personalized learning pathways, which have been shown to improve student engagement and academic performance across diverse educational settings \cite{RuenMeylani2024, PradeepBMane2024}. For instance, AI literacy apps that adjust text complexity and offer real-time decoding support have been particularly effective for students with reading difficulties, leading to notable improvements in reading fluency compared to traditional instruction methods \cite{HasanSari2024}. This personalized approach not only helps low-performing students catch up more quickly but also provides high-performing students with advanced challenges, thereby promoting an inclusive learning environment that caters to a wide range of abilities \cite{StephenBezzina2024}. The integration of AI in education has also been associated with increased student engagement and satisfaction, as evidenced by improvements in GPA and reduced study hours \cite{BenWard2024}. Moreover, AI-enhanced educational platforms have been shown to boost student performance by up to 30\% and enhance engagement by over 60\%, highlighting the transformative potential of AI in education \cite{PradeepBMane2024}. However, challenges such as data quality, ethical considerations, and the need for educator support remain critical for the successful implementation of these systems \cite{RuenMeylani2024, WaiYieLeong2024}. Addressing these challenges is essential to fully realize the benefits of AI-driven adaptive learning systems, which have the potential to create more equitable and effective educational environments by providing continuous, targeted support to all students \cite{HasanSari2024, WaiYieLeong2024}.

AI's integration into special education has significantly enhanced functional and social outcomes for students with disabilities, particularly in fostering inclusion and improving communication skills. AI technologies, such as AI-enhanced communication devices, have been instrumental in supporting non-verbal students by improving their expressive communication and autonomy. These devices, when optimized with AI for speed and predictive capabilities, enable students to engage more actively in dialogues, thereby enhancing their language abilities and social integration over time \cite{AndreaRHarkinsBrown2025, MunikrishnaiahSundaraRamaiah2024}. For students with Autism Spectrum Disorder (ASD), AI interventions have shown promise in addressing social communication challenges, as these students often find technology's predictability and limited social demands appealing \cite{SofiaKotsi2025}. AI-powered tools, such as speech recognition and text-to-speech systems, have also been pivotal in increasing accessibility and supporting academic success, thereby promoting social justice and equity in education \cite{ErenKamber2025}. Furthermore, AI's role in adaptive learning environments has been shown to significantly improve academic performance, engagement, and social interaction among students with special needs in inclusive classrooms \cite{SYARIFMAULIDIN2024}. Teachers have observed that AI tools, such as AI avatars for practicing presentations, help students gain confidence and initiate interactions, which are crucial for social integration \cite{AndreaRHarkinsBrown2025}. The personalized nature of AI-driven educational tools allows for tailored learning experiences that cater to individual needs, thereby fostering a more inclusive environment where all students, regardless of their disabilities, can participate meaningfully in classroom activities \cite{MBhakiyasri2024, SilvioMarcelloPagliara2024}. However, while AI offers substantial benefits, it is essential to address ethical considerations, such as privacy and potential biases, to ensure these technologies are implemented responsibly and effectively \cite{AndreaRHarkinsBrown2025, SilvioMarcelloPagliara2024}.

AI technologies, such as adaptive learning systems and intelligent tutoring systems, enable educators to personalize learning experiences by identifying individual student needs and tailoring interventions accordingly \cite{SMahalakshmi2025, MeltemTakn2025}. This personalized approach is particularly beneficial for students with disabilities, as AI can adapt educational content to meet their unique requirements, thereby improving accessibility and engagement \cite{SMahalakshmi2025, ShaleceKohnke2025}. Moreover, AI analytics provide educators with insights into student performance trends, allowing them to identify which students may be struggling and require additional support. This capability facilitates timely interventions, reducing the likelihood of students falling behind and ensuring that resources are allocated effectively to support diverse learning needs \cite{WaiYieLeong2025, MarkTreve2024}. For instance, AI-driven progress monitoring can reveal that certain interventions, like a phonics app, are less effective for English learners, prompting schools to supplement these with additional language support \cite{MarkTreve2024}. Furthermore, AI's role in special education is transformative, as it helps in accurately identifying students who need formal services, thus reducing inappropriate special education placements and ensuring that students receive the right level of support at the right time \cite{AndreaRHarkinsBrown2025}. The use of AI in education also extends to optimizing administrative tasks and enhancing pedagogical practices, which further supports inclusive education by freeing up educators to focus more on student engagement and personalized instruction \cite{SumartonoSumartono2025, NazarenaPatrizi2025}. However, the integration of AI must be approached with caution, considering ethical implications such as data privacy and algorithmic bias, to ensure that AI systems promote equity rather than exacerbate existing disparities \cite{WanChongChoi2025, ShaleceKohnke2025}.

Table \ref{tab:aispecialeducation} mentions all the key outcomes of AI integration in special education and language barriers.

\begin{table}[h]
    \centering
    \caption{Key Outcomes of AI Integration in Special Education and Language Barriers}
    \begin{tabular}{p{4cm} p{8cm} p{2cm}}
        \hline
        \textbf{Outcome} & \textbf{Description} & \textbf{Reference} \\
        \hline
        Personalized Learning Experiences & Adaptive learning systems adjust content based on individual student needs. & \cite{SYARIFMAULIDIN2024, WasimAhmad2024, SilvanaMariaAparecidaVianaSantos2024} \\

        Improved Academic Performance & AI-based learning systems enhance academic scores and engagement. & \cite{SYARIFMAULIDIN2024, WasimAhmad2024, ElizabethAngoFomusoEkellem2024} \\
    
        Enhanced Engagement and Social Interaction & AI tools foster participation and collaboration in inclusive classrooms. & \cite{WasimAhmad2024, AliaElNaggar2024, ElizabethAngoFomusoEkellem2024} \\
     
        Assistive Technologies for Disabilities & AI enables interaction for students with physical or cognitive disabilities. & \cite{PandiC2023, WasimAhmad2024, MohsenMahmoudiDehaki2024} \\
 
        Speech and Language Development & AI supports language acquisition for intellectually disabled children. & \cite{HanadiHussainAlHadiAlQahtani2024, MohsenMahmoudiDehaki2024} \\
    
        Addressing Language Barriers & AI tools facilitate real-time translation and communication in multilingual classrooms. & \cite{KaloyanDamyanov2024, ElizabethAngoFomusoEkellem2024} \\
  
        Cultural Sensitivity and Inclusivity & AI adapts educational content to diverse cultural backgrounds. & \cite{ElizabethAngoFomusoEkellem2024} \\
     
        Ethical Considerations and Challenges & Ensuring equitable access and addressing biases in AI algorithms. & \cite{SilvanaMariaAparecidaVianaSantos2024, AliaElNaggar2024} \\
        \hline
    \end{tabular}
    \label{tab:aispecialeducation}
\end{table}

\section{Recommendations}

\begin{figure}[h]
    \centering
    \includegraphics[width=0.8\linewidth]{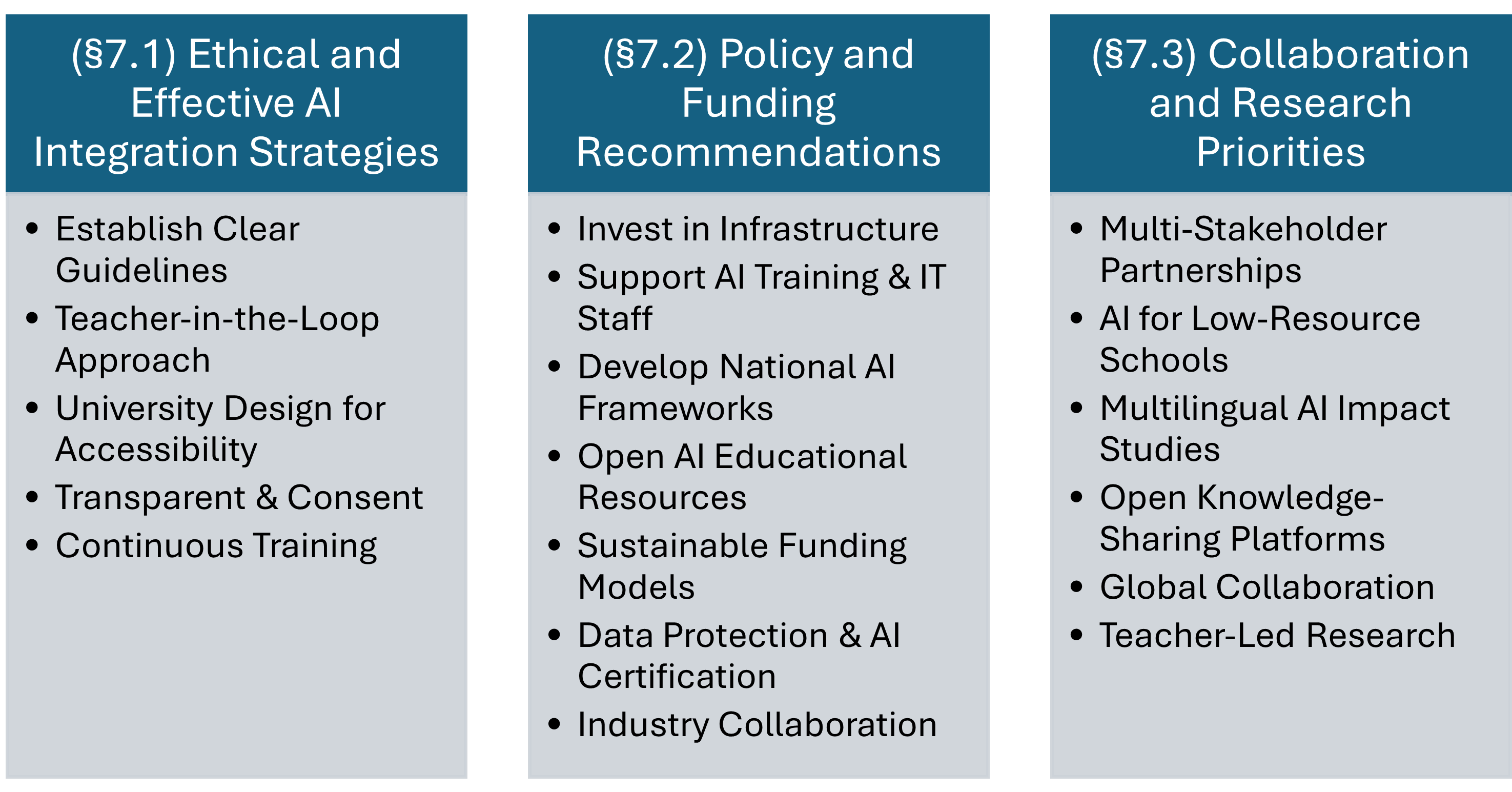}
    \caption{Recommendations to responsible AI implementation.}
    \label{fig3}
\end{figure}

Some key recommendations have been identified to ensure the ethical and effective integration of AI in inclusive education. Figure \ref{fig3} summarizes these recommendations for a structured approach to responsible AI implementation. 

\subsection{Ethical and Effective AI Integration Strategies}

To harness AI for inclusive education responsibly, the author recommends several strategies centered on ethical use and thoughtful integration. First, it is crucial to establish clear guidelines and policies at the school or district level for AI use in classrooms. These should outline acceptable uses of AI, data privacy protections, and processes for monitoring algorithmic fairness. For example, schools should adopt policies ensuring that any AI system handling student data complies with privacy laws and undergoes regular audits for bias. An AI Ethics Committee or working group can be formed within the district, including educators, parents, and IT experts, to evaluate AI tools before adoption and to continuously review their impact. This committee would help vet whether an AI reading app has been tested with English Language Learners (ELLs) and special needs students, or whether a learning analytics platform provides interpretable and fair metrics. By having such oversight, schools can prevent problematic AI implementations and build trust among stakeholders.

Another key strategy is to ensure human-in-the-loop at all critical junctures of AI usage. Teachers and support staff should retain ultimate control over educational decisions, with AI serving as an advisor or assistant. This means, for instance, AI-generated recommendations (be it grouping suggestions, personalized content, or flags of student risk) should be presented transparently to teachers, who then confirm or adjust them based on their professional judgment. AI systems must be designed to explain their suggestions in user-friendly terms  a concept known as explainable AI. Vendors should be encouraged to provide interfaces that show why the AI is recommending a certain exercise for a student (e.g., Student struggled with two-digit addition in the last five attempts, so I am providing more practice in that area) rather than just pushing content mysteriously. This not only aids teacher understanding but also allows them to catch potential AI errors or biases. In line with this, teachers should always have the ability to override AI decisions. For example, if a teacher knows a student is having an off day due to personal issues, they might override an AIs suggestion to give a test and postpone it. By keeping educators in the loop, the author guards against over-reliance on algorithms and maintains the primacy of human-centric education.

The author also recommends adopting universal design principles in AI tool deployment. This means choosing and configuring AI tools to be accessible to the widest range of students from the outset. Features like adjustable reading speeds, multilingual support, captioning and audio description in multimedia, and alternative input methods (voice, touch, switch devices for those who cannot use standard interfaces) should be enabled wherever possible. If an AI platform does not natively support such accessibility, educators should work with the vendor or use assistive tech add-ons. For example, if a math AI tutorial is heavily visual, ensure a screen reader can interpret it for blind students or provide a tactile printout alternative. Ethically, no student should be excluded from the benefits of an AI-enhanced activity due to a disability. When piloting new AI tools, involve students with diverse needs in user testing and gather their feedback on usability and comfort. Their insights can drive necessary modifications or determine if a tool is unsuitable.

To further ethical integration, its also recommended to implement transparency and consent practices. Parents should be informed about what AI systems are being used, what data is being collected, and how it benefits learning. Schools can hold informational sessions or send out easy-to-understand briefs on their AI initiatives, addressing questions and concerns. In sensitive cases (like AI that monitors behavior or uses video), obtaining opt-in consent is advised. Alternatively, providing equivalent non-AI pathways for families not comfortable with AI ensures no one is forced into a situation against their values. Transparency can alleviate fear  for instance, explaining that a writing AI checks grammar and provides suggestions can help parents see it as a learning aid rather than cheating or surveillance. When issues do arise (like an AI misinterpreting something due to bias), openly communicating these incidents and steps taken to fix them will enhance an ethical place.

Finally, continuous professional development in ethical AI usage is needed. Training sessions should not just cover how to use AI tools, but also ethical scenarios and decision-making. Teachers might be presented with case studies (e.g., an AI flags only minority students as off-task  how to respond?) and discuss appropriate actions. This prepares teachers to handle the nuanced dilemmas that can come with AI recommendations or data. It also empowers them to be advocates; a teacher trained to notice bias can alert the school or vendor to improve the system, thus contributing to wider ethical improvement.

\subsection{Policy and Funding Recommendations}
Policymakers and educational leaders play a critical role in scaling AI for inclusive education equitably. One major recommendation is to invest in infrastructure and access such that every school has the baseline technology needed for AI tools. Governments should consider programs that fund high-speed internet for schools, provide grants for purchasing devices (laptops, tablets) in sufficient quantities, and maintain up-to-date hardware. Just as electricity and heating are considered basic utilities for schools, digital infrastructure must be seen as essential. Some regions have initiated 1:1 device programs or invested in community internet access points; these efforts should explicitly be linked to enabling advanced learning technologies like AI, and extended to early childhood centers and special education units which are sometimes overlooked. Alongside infrastructure, funding must support tech support and training. Its not enough to drop devices into schools; consistent technical support (IT staff) and budget for teacher training on AI integration are needed. Governments could earmark a portion of education budgets or special innovation funds for annual teacher workshops on educational AI, partnerships with universities for teacher upskilling, or for hiring additional support specialists who can coach teachers in classrooms.

At a higher level, the author recommends that national or state education authorities develop a comprehensive AI-in-education strategy or framework that includes inclusion as a guiding principle. This means any official roadmap for adopting AI in schools should highlight ensuring accessibility and equity as core goals, not afterthoughts. For instance, if a Ministry of Education launches an AI tutoring program, it should concurrently launch initiatives to reach rural schools, to localize the AI in multiple languages, and to include content for special education. One actionable item is the creation of open educational AI resources. Governments or philanthropic organizations could sponsor the development of AI-driven learning platforms as open-source or public goods, focusing on areas where commercial incentives are weaker (like less common languages or specialized tools for low-incidence disabilities). By doing so, they reduce dependence on expensive proprietary systems and can adapt tools to local curricula and values. For example, an AI literacy platform built as an open project could be adapted by different countries or states to include culturally relevant stories and support various languages at a fraction of the cost of each buying separate solutions. Policy can encourage this through funding challenges or public-private partnerships.

Funding should also support further research and pilot programs specifically in inclusive applications of AI. This includes longitudinal studies to see how AI impacts diverse learners over time, and pilot projects in high-need communities to innovate new uses (like AI for indigenous language education, or AI for early screening of learning difficulties). The findings from such research should inform policy updates regularly  for instance, if evidence shows AI tutoring greatly helps in numeracy for special needs students, governments could mandate or encourage its use in all special education programs, with funding attached to adoption and training. Conversely, research might point out pitfalls, enabling policy to set necessary regulations (e.g., limiting use of AI during certain formative periods to ensure basic skill development is not impeded).

Another recommendation is around funding models that ensure sustainability and equity. Rather than one-off grants that create a burst of tech adoption followed by stagnation (when equipment becomes outdated or subscriptions lapse), funding should be structured for long-term support. This could mean recurring grants or matching funds to maintain and upgrade AI tools, especially in lower-income districts. Additionally, consider central procurement at the state/national level of key AI platforms to negotiate better pricing and uniform access  similar to how textbooks are often purchased. When deploying specialized AI (like for special needs), cover the extra costs so that schools with higher numbers of such students are not financially strained (for example, voice recognition for speech impairments might need specialized software or hardware). Equity-focused funding might also allocate more resources to schools that have catching up to do in tech capacity.

On the policy front, data governance and protection laws must evolve in parallel. Governments should articulate how student data used in AI is protected, who owns algorithmic outputs, and possibly provide certification for AI tools that meet high standards of privacy and non-bias. This is analogous to how medical devices are regulated  educational AI that influences learning paths might require a certification mark that indicates its been evaluated for effectiveness and fairness. While this could be complex, starting with guidelines or voluntary certification could move the industry in the right direction.

Finally, collaboration between the education sector and tech industry should be fostered by policy. Incentivize companies to work on inclusive AI solutions (for example, tax incentives or recognition for developing products for under-served languages or disabilities). Encourage contests or innovation labs focusing on AI for inclusive education, perhaps hosted by education ministries or international bodies. The goal is to steer the immense innovation energy of the tech industry toward solving the most pressing inclusion challenges.

\subsection{Collaboration and Research Priorities}
To truly realize the potential of AI in fostering inclusive education, a collaborative ecosystem is needed, bridging educators, technologists, researchers, and communities. One key recommendation is to establish multi-stakeholder partnerships when designing or deploying AI solutions. For example, a school district planning to roll out an AI-based learning platform could partner with a universitys education department to co-design the implementation and evaluation, with input from teachers and even students. In such a partnership, teachers contribute practical insights (like what kind of feedback is most useful), researchers ensure rigorous evaluation and adapt the theoretical framework, and technologists from the AI vendor adjust the product to better meet educational needs. A concrete outcome of such collaboration could be co-created lesson plans that integrate the AI tool seamlessly, or an improvement in the AI interface based on teacher suggestions. Community involvement is also important; parents or advocacy groups for students with disabilities can provide perspective on any concerns and help tailor the communication around AI use. When stakeholders move in step rather than in isolation, the chance of success and acceptance of AI initiatives is much higher.

Another recommendation is to prioritize certain research areas that can significantly advance inclusive AI use. One priority is AI for low-resource contexts: research that focuses on how to make AI tools effective even in schools with limited connectivity or devices. This could involve developing offline-capable AI systems, or AI that can run on cheaper hardware like Raspberry Pi-based devices. Such research directly addresses equity, aiming to not leave rural or underfunded schools behind. Similarly, research on multilingual AI education is critical  making AI truly language-inclusive beyond English and other widely used languages. This includes algorithms that can learn from relatively small datasets in a local language, or systems that can switch between languages fluidly to support bilingual education. There is a notable gap here that academic researchers in NLP (Natural Language Processing) and education could collaborate on, potentially supported by government grants or international funds.

Another research priority is understanding long-term impacts of AI-assisted learning on different student populations. For inclusion, one needs to know: Do benefits sustain over time? How does early exposure to AI tutors affect a childs later independent learning skills? Are there any negative side effects (for instance, reduced social interaction or problem-solving skills if over-relying on AI)? Longitudinal studies that follow cohorts of students who extensively use AI supports versus those who dont could provide answers, guiding how one blends AI into pedagogy without undermining other developmental aspects. Its important these studies specifically examine subgroupsstudents with disabilities, ELLs, etc.to ensure AI is serving them well in the long run. Research should also probe the psychological aspect: how do students perceive learning with AI? Does it change their self-efficacy or mindset? This ties into making sure AI use promotes positive attitudes and does not, for instance, make students overly reliant or reduce their perseverance (an issue some educators fear).

The author also recommends open knowledge-sharing platforms for educators to share experiences, lesson plans, and data from their AI integration efforts. This could be something like a global repository or community of practice (similar to how teachers share resources on sites like TES or TeachersPayTeachers, but focused on AI tools). Within this, teachers can share what worked or didnt for inclusionfor example, a teacher might upload a case study of using a speech AI with a student with apraxia, detailing the adjustments made. These real-world notes are invaluable research fodder and practical guidance for peers. Educational researchers or organizations (like ISTE or UNESCO) could moderate and curate these platforms, possibly combining them with formal research findings for a rich knowledge base.

Collaboration is also needed between countries and regions. High-income countries might lead in AI ed-tech, but low-income countries have a wealth of knowledge in doing inclusion with minimal resources. A global collaboration could pair schools or districts across different contexts to pilot AI tools and share outcomes. For instance, an inclusive school in Finland might partner with one in India to both use a particular AI math tutor, then share how it performed in their contexts. This cross-pollination would surface different challenges and creative solutions, informing better design universally.

Finally, empowering teachers as researchers in their own classrooms is a valuable collaborative approach. Teachers can be trained in action research methodology to systematically track how AI affects their students and to experiment with interventions. If they document and share this, it adds to collective knowledge. Some districts encourage teacher-led inquiry groups (PLCs) focusing on technology integrationthese could be formalized with a connection to academic mentors so that teachers findings feed into broader research and vice versa.

\subsection*{Implementation Roadmap for Educators}

\begin{enumerate}
    \item \textbf{Assess Needs:} Identify students who face language or accessibility barriers.
    \item \textbf{Select Tools:} Choose AI applications appropriate for the classroom context.
    \item \textbf{Train Staff:} Offer brief workshops to educators on ethical and effective AI use.
    \item \textbf{Pilot and Monitor:} Start with small-scale pilots and gather feedback.
    \item \textbf{Scale and Sustain:} Expand use after evaluation and refine based on student progress.
\end{enumerate}

\section{Conclusion}
The integration of artificial intelligence into early inclusive education holds remarkable promise, but realizing its benefits universally will require deliberate effort, ethical vigilance, and sustained collaboration. This chapter has examined how AI can support young learners by breaking language barriers through real-time translation, by personalizing learning experiences for students with special needs, and by augmenting teacher capacities to better meet diverse needs. It has been seen that AI-driven tools  from adaptive tutors to assistive communication devices  can increase student engagement, accelerate skill acquisition, and empower previously marginalized students to participate more fully in learning. Early case studies and implementations show encouraging outcomes, such as non-verbal children finding their voice via predictive AAC or multilingual students keeping pace with their peers thanks to translation support. These successes illustrate AIs potential to serve as a powerful ally in the quest for truly inclusive, equitable education.

At the same time, it can be made clear that these technologies are not a panacea and come with their own set of challenges. Issues of equitable access, such as the digital divide and bias in AI systems, pose real risks of widening gaps if not addressed head-on. It is imperative that as educators and policymakers, one ensures that AI tools are accessible and effective for all students, including those in under-resourced settings and those with atypical needs. This means investing in infrastructure, choosing or developing AI solutions that are culturally and linguistically inclusive, and maintaining humans in the loop to catch and correct the limitations of algorithms. Teachers will remain irreplaceable as facilitators, decision-makers, and emotional supporters in the classroom; AI should free them from drudgery, not from their connection with students. The ethical use of AI  safeguarding privacy, ensuring transparency, and preventing any form of discrimination  must be a foundational principle. The recommendations provided in this chapter, from instituting clear AI governance policies to fostering collaborations for inclusive design, are intended as guideposts to navigate these AI topics responsibly.

The journey of integrating AI in education is just starting, and future research will play a critical role in guiding its trajectory. The author calls for longitudinal studies and continued experimentation to deepen the understanding of how different AI interventions impact learning outcomes and inclusion over time. For example, a future research direction is to track a cohort of students with special needs who use AI supports throughout their schooling to see how it affects their academic progression and social inclusion into adulthood. Another avenue is researching student agency with AI: how can students eventually take the reins, perhaps even training or tweaking their own AI helpers, thereby learning meta-cognitive and technical skills in tandem? Additionally, as AI technology evolves (with advancements like more emotionally perceptive AI or sophisticated speech and gesture recognition), investigating how these can be harnessed for education without infringing on student development or privacy will be important. The author also encourages cross-disciplinary studies  educators teaming up with computer scientists, ethicists, speech therapists, etc.  to innovate holistic solutions. There is much to learn about optimal teacher-AI collaboration models, effective training approaches, and the psychological impacts of learning with AI.

As a final analysis, the advent of AI in early education is nothing else but a tool  a very powerful one  that reflects the intentions and frameworks in which it is employed. Used wisely, AI can help fulfill the long-standing dream of inclusive education by tailoring learning to each childs needs and bridging gaps that human resources alone have struggled to address. It can act as a force multiplier for teachers, amplifying their ability to reach every learner, and as a personalized scaffold that gives every student a boost where and when they need it. However, if used carelessly or inequitably, the same tool could reinforce biases or leave some children further behind. The call to action, therefore, is for all stakeholders in education to engage proactively with this technology: to shape it, guide it, and critique it, with the unwavering goal that it serves the best interests of every child. The future of AI in inclusive education will be what can be collectively made of it. Let us make it one where technology and humanity coalesce to extend the reach of quality education to all learners, celebrating diversity and enabling each child to thrive. With thoughtful integration and continued advocacy for equity, AI can be a catalyst that helps transform inclusive education from ideal to practical reality, ensuring no learner is left behind in the AI age.

\end{document}